\documentclass[12pt,preprint,iop]{emulateapj}

\usepackage{epsfig, natbib} 
\usepackage{mathrsfs,amssymb}
\usepackage{graphicx,latexsym}
\usepackage{ulem}
\usepackage{enumitem}
\usepackage[]{subfigure}
\setenumerate[1]{label=\Roman*.}
\setenumerate[2]{label=\Alph*.}
\setenumerate[3]{label=\arabic*.}
\setenumerate[4]{label=\roman*.}
\setenumerate[5]{label=\alph*.}

\newcommand{\etal}{{\rm et~al.\/}}
\newcommand{\fesc}{\mbox{$f_{\rm esc}$}}

\newcommand{\Hline}[1]{\mbox{H{\footnotesize {#1}}}}
\newcommand{\Ha}{\Hline{\mbox{$\alpha$}}}

\newcommand{\HI}{{\sc \rm HI}}

\newcommand{\hii}{H{\sc ii}}
\newcommand{\iraf}{{\sl \rm IRAF\/}}

\newcommand{\Msun}{\mbox{${\cal M}_\odot$}}

\newcommand\na{NewA}

\newcommand{\nii}{[N{\sc ii}]}
\newcommand{\oiii}{[O{\sc iii}]}
\newcommand{\oii}{[O{\sc ii}]}

\newcommand{\Qo}{\mbox{$Q_0$}}

\newcommand{\SigSFR}{\mbox{$\Sigma_{\rm SFR}$}}
\newcommand{\sii}{[S{\sc ii}]}
\newcommand{\siii}{[S{\sc iii}]}

\slugcomment{*** Resubmitted 27 October 2013 ***}

\shorttitle{Ionization-Parameter Mapping with MMTF}
\shortauthors{Zastrow \etal}

\begin{document}

\title{New Constraints on the Escape of Ionizing Photons from Starburst Galaxies Using Ionization-Parameter Mapping}

\author{Jordan Zastrow \altaffilmark{1},
        M.S.\ Oey\altaffilmark{1},
        Sylvain Veilleux\altaffilmark{2},
        Michael McDonald\altaffilmark{3}}

\altaffiltext{1}{Department of Astronomy, University of Michigan, 500 Church Street, 830 Dennison, Ann Arbor, MI 48109-1042, $jazast@umich.edu$}
\altaffiltext{2}{Department of Astronomy, University of Maryland, College Park, MD 20742, USA}
\altaffiltext{3}{Kavli Institute for Astrophysics and Space Research, MIT, Cambridge, MA 02139, USA}

\begin{abstract}

The fate of ionizing radiation in starburst galaxies is key to understanding  cosmic reionization. However, the galactic parameters on which the escape fraction of ionizing radiation depend are not well understood. Ionization-parameter mapping provides a simple, yet effective, way to study the radiative transfer in starburst galaxies. We obtain emission-line ratio maps of \siii/\sii\ for six, nearby, dwarf starbursts: NGC~178, NGC~1482, NGC~1705, NGC~3125, NGC~7126, and He~2-10.  The narrow-band images are obtained with the Maryland-Magellan Tunable Filter at Las Campanas Observatory. Using these data, we previously reported the discovery of an optically thin ionization cone in NGC 5253, and here we also discover a similar ionization cone in NGC 3125. This latter cone has an opening angle of $40\pm5\degr$ (0.4 ster), indicating that the passageways through which ionizing radiation may travel correspond to a small solid angle.  Additionally, there are three sample galaxies that have winds and/or superbubble activity, which should be conducive to escaping radiation, yet they are optically thick. These results support the scenario that an orientation bias limits our ability to directly detect escaping Lyman continuum in many starburst galaxies. A comparison of the star-formation properties and histories of the optically thin and thick galaxies is consistent with the model that  high escape fractions are limited to galaxies that are old enough ($\gtrsim$ 3 Myr) for mechanical feedback to have cleared optically thin passageways in the ISM, but young enough ($\lesssim$ 5 Myr) that the ionizing 
stars are still present.

\end{abstract}


\section{Introduction} \label{s:intro}

Between $z\sim11$ and $z\sim6$, the universe experienced a change of state known as cosmic reionization \citep[e.g.,][]{b:Shull_apj12,b:Fan_araa06,b:Fan_aj01}.  The source for cosmic reionization is still unclear; the primary contenders are star-forming galaxies and quasars.  However, the space density of quasars peaks at $z\sim3$ and drops below the threshold needed to supply the required amount of ionizing radiation at $z\gtrsim 6$ \citep{b:Madau_apj99,b:Fan_aj01,b:Meiksin_mnras05, b:Shankar_apj07}.  The massive stars found in high-redshift star-forming galaxies have the potential to provide enough ionizing radiation, but whether they contribute significantly to the intergalactic ionizing emissivity is not yet established \citep[e.g.,][]{b:Yajima_arXiv12}. In order to account for cosmic reionization, the mean fraction of ionizing radiation that escapes the galaxy population $<\fesc>$, needs to be at least 20\% \citep[e.g.,][]{b:Bouwens_apj10}.  

In theory, starburst galaxies should be porous to ionizing radiation \citep[e.g.,][]{b:Clarke_mnras02,b:Paardekooper_aap11}.  Stellar winds and supernovae, in conjunction with radiation pressure, will punch holes in the interstellar medium (ISM) \citep[e.g.,][]{b:MacLow_apj88}, and thereby create low density passageways through which ionizing radiation may escape \citep[e.g.,][]{b:Dove_apj00}.  Feedback from massive stars is strongest in galaxies with high star-formation rates, making starbursts attractive candidates for the source of cosmic reionization. 

The observational evidence for significant \fesc\ at high redshift is
inconclusive. Using a sample of 29 Lyman break galaxies
(LBG), \citet{b:Steidel_apj01} showed that at least some $z\sim3$
starbursts have large escape fractions.  However, Lyman continuum narrow-band imaging
studies with larger galaxy samples found escaping Lyman continuum in
only 10\% of the galaxies observed
\citep[e.g.,][]{b:Shapley_apj06,b:Iwata_apj09}. Observations by
\citet{b:Nestor_apj11}, put the mean $z\sim 3$ escape fraction at
$<\fesc>\sim 0.12$.  Is it important to note that \citet{b:Vanzella_apj12} recently
illustrated that a majority of narrowband Lyman continuum detections
could actually be low $z$ interlopers, which would drive the
observational \fesc\ even lower.   

At intermediate to low redshift, some evidence exists for significant \fesc\ from LGB analogs. A subset of these galaxies with extreme feedback have favorable conditions for escaping Lyman continuum, such as low opacity in the ISM \citep{b:Heckman_apj11}. However, the fraction of galaxies with these conditions is still small, indicating a low mean \fesc\ \citep[e.g.,][]{b:Siana_apj10}.  Looking at $z\sim0$ galaxies, the detection rate drops further, with only two examples of local starbursts that have measurable \fesc\ \citep{b:Leitet_aap13, b:Leitet_aap11}.  Most other studies find no direct evidence for escaping Lyman continuum in local starbursts \citep{b:Bridge_apj10,b:Grimes_apjs09,b:Heckman_apj01,b:Leitherer_apj95}. 

In order to determine the mean galactic \fesc\ from observed escape
fractions, we need to disentangle the effects of observational biases
from trends with intrinsic galaxy properties.  In the massive-star
feedback-driven model, the escape fraction will depend on the star
formation rate, the ISM morphology and density \citep{b:Fernandez_apj11,b:Benson_arXiv12}, 
and distribution of star formation.  However, the winds that clear the ISM are not isotropic, which leads to a preferred direction for escape and finite opening angles \citep{b:Veilleux_araa05}. These effects create observational dependencies with the orientation of the host galaxy.  Finally, measuring the Lyman continuum at any redshift is a challenge.  Most of the flux is absorbed by the intergalactic medium, thus the signal is low, which requires long integration times. For targets at higher redshift, low-redshift interlopers confuse the statistics \citep{b:Vanzella_apj12}. Some studies measure the optical depth to UV lines in the ISM in lieu of direct measurements of the Lyman continuum. They obtain the escape fraction after making assumptions about the gas distribution and the relative optical depths in the Lyman continuum and the observed ion \citep[e.g.,][]{b:Heckman_apj11}. 
 
In this study we use a method that is complementary to other
  established approaches for investigating the factors that affect the
  escape of ionizing radiation from starbursts.  We map the ionization
  structure of extended ionized gas in six local starbursts using
  emission-line ratio images. Spatial changes in emission-line ratios
  reveal the passage of ionizing radiation through the galaxy \citep{b:Pellegrini_apj12}.
  Thus, by studying the ionization structure of extended ionized gas in our sample, we can determine the optical depth of the galaxies and evaluate the likelihood of escaping Lyman continuum. 
\section{Method and Data} \label{s:methdata}

\subsection{Method}\label{s:method}

Determining the fate of ionizing radiation in starburst galaxies is a challenging observational problem. Here, we present an approach that uses the technique of ionization parameter mapping \citep[IPM;][]{b:Pellegrini_apj12}.  Spatial changes to the ionization parameter, which evaluates the ionizing photon density relative to the gas density, trace the passage of ionizing radiation through the ISM.  In order to track these spatial changes, we create maps of the starbursts using the \siii/\sii\ line ratio, which acts as a proxy for the ionization parameter.  Optically thin and thick regions exhibit different ionization parameter morphologies.  In regions that are optically thick, there is a transition from high to low ionization states at the interface of the ionized region and the neutral environment, whereas optically thin regions lack this transition zone \citep{b:Pellegrini_apj12}. Thus, by using ratio maps of two ions, one can distinguish between optically thick and thin regions \citep{b:Pellegrini_apj12}.  Across an entire galaxy, one can thus use the ionization structure to evaluate whether and in what manner ionizing radiation escapes the galaxy \citep[e.g.,][]{b:Zastrow_apj11,b:Pellegrini_apj12}.  This method has the advantage that it can be accomplished with narrow-band ground-based imaging, which makes it more observationally straightforward than other methods.  
 
\subsection{Data}\label{s:data}

The galaxies selected for this study are all nearby dwarf
starbursts. They have star formation rates (SFR) of 0.05--5 \Msun/yr and stellar
masses log(M$_*$) $\sim 8-10$.  Since massive star feedback is thought
to play a key role in the escape of Lyman continuum (LyC) 
photons
\citep{b:Heckman_apj11}, we select galaxies that have
evidence in the literature for winds and expanding
bubbles. Table 1 shows the properties of our sample.  We calculate the stellar masses  (Column 2) using the $K$-band magnitudes from \citet{b:Skrutskie_cat03} and the mass to light ($M$/$L$) ratios following \citet{b:Bell_apjs03}, which use the galaxies' $B-V$ or $B-R$ colors to derive $M/L$.  We
  restrict this study to starbursts within 45 Mpc, which corresponds to spatial scales of $\sim 200$ pc/arcsec, to ensure
high enough spatial resolution to trace the ionization
structure of the extended gas. 

We obtained narrow-band emission-line imaging in \Ha, \sii$\lambda6716$, and \siii$\lambda9069$\ on the nights of 2009 July 7--11 and 2010 February 12--14.  The wavelengths quoted above are the rest frame wavelengths of the emission lines.  We imaged the galaxies using the Maryland-Magellan Tunable Filter \citep[MMTF;][]{b:Veilleux_aj10}, which is mounted on the Inamori-Magellan Areal Camera and Spectrograph (IMACS) on the Magellan Baade telescope at Las Campanas Observatory.  For every emission-line image, we obtained paired continuum exposures of the galaxies.  Our observing log is presented  in Table \ref{t:obs}.  For each galaxy (Column 1), we list the targeted emission line and central bandpass (Columns 2, 3), exposure times (Columns 4), and the 3$\sigma$ surface brightness limit in the reduced image (Columns 5).

We reduced the MMTF data using version 1.4 of the MMTF data reduction pipeline\footnote{http://www.astro.umd.edu/\textasciitilde veilleux/mmtf/datared.html}. This pipeline performs bias and flatfield corrections and then subtracts the sky using an azimuthal averaging. Next, it corrects cosmic rays and bad CCD pixels before mosaicking the different CCD chips into one image.  Finally, it averages the individual emission-line and continuum images.  We flux calibrated the images using standard stars LTT 1020, 6248 and 7987 \citep{b:Hamuy_pasp94}.  We note that for the February 2010 observing run, our measured \Ha\ fluxes are a factor of 2--3 lower than the \Ha\ fluxes reported in the literature.  A combination of shallower exposure times, combined with errors in the velocity offsets of our central bandpass, wind activity that shifts flux out the narrow MMTF bandpass, and uncertainty in the \nii\ correction factors for the literature values, lead to this offset.  However, a comparison with long-slit spectra confirms that the line ratios are sound.

In order to cleanly subtract the continuum from the emission-line images, we first aligned the emission-line and continuum frames using \iraf\footnote{\iraf\ is distributed by NOAO, which is operated by AURA, Inc., under cooperative agreement with the National Science Foundation.} tasks {\tt wregister} and {\tt imalign}.  Next, we used {\tt psfmatch} to convolve the image with the better seeing with the psf from the other. Finally, we used several bright stars in the field and the task {\tt mscimatch} to determine the proper scaling to match counts in the stars between the images.  This last step was performed iteratively until we obtained a clean continuum subtraction.  We present the final continuum-subtracted \Ha\ images and three-color composite images of the galaxies in our sample in Section \ref{s:results}. 

Scattered light in the instrument causes an optical artifact that is
not easily corrected for and that stretches across CCD chips 3 and 5.  For
the most part, we avoid this issue by placing our targets on the other
chips. However, for NGC 178, most of our observations have the galaxy
sitting on chips 3 or 5.  To mitigate the artifact, we mask the affected region in the flat field.  While this increases the flux error on this galaxy, this effect is much less significant than the alteration in flux caused by the artifact on the flat. 

We also note that the data from the 2009 July observing run are bright time observations.  As a result of scattered light in the f/2 camera and the MMTF instrument, these observations are challenging to reduce.  The emission-line images of NGC~178 and NGC~7126 show many background sky features that are caused by these effects (Figure \ref{f:img_n178} and \ref{f:img_n7126}).

{ 
\begin{deluxetable}{l c  c c c }
  \tablewidth{0pt}
  \tabletypesize{\footnotesize}
  \tablecaption{Sample\label{t:props}}
  \tablehead{\colhead{Galaxy} &
             \colhead{log($M_*$)$^a$} &
             \colhead{Distance} &
             \colhead{$L(\Ha)$} &            
             \colhead{Ref$^b$}\\
             \colhead{  } &
             \colhead{[\Msun]} &
             \colhead{[Mpc]} &   
             \colhead{[erg s${-1}$]} &        
             \colhead{  }}
\startdata
NGC~178 & 9.24 & 20.6 & $6.1\times 10^{40}$&(5),(8)\\
NGC~1482& 10.56 & 22.6 & $1.12\times 10^{42}$ &(2)\\
NGC~1705& 8.37 & 5.1 & $8.09\times 10^{39}$ &(3),(7) \\ 
NGC~3125& 9.09 & 11.5 & $4.49\times 10^{40}$&(1),(7)\\ 
NGC~5253& 9.05 & 3.8 & $3.71\times 10^{40}$ & (9),(7)    \\
NGC~7126& 10.44  & 45.5 & $6.31\times10^{41}$ & (5),(8) \\
He~2-10 & 9.50 & 9 & $6.09\times 10^{40}$&(6),(4) 
\enddata
\tablenotetext{a}{Stellar masses calculated using the $M/L$ ratios derived from \citep{b:Bell_apjs03}, 2MASS $K$ magnitudes \citep{b:Skrutskie_cat03}, and optical colors as follows: B-R colors from \citet{b:GildePaz_apjs03} for NGC~1705 and NGC~3125, B-V colors from \citet{b:Moustakas_apjs10} for NGC~1482, B-V colors from \citet{b:deVaucouleurs_rc391} for NGC~7126 and NGC~178, B-V colors from \citet{b:Taylor_apj05} for NGC 5253, and B-V colors from \citet{b:Ho_apjs11} for He~2-10.  }
\tablenotetext{b}{References listed for D and $L(\Ha)$:(1) \citet{b:Schaerer_aaps99} (2) \citet{b:Kennicutt_pasp11} (3) \citet{b:Tosi_aj01} (4) \citet{b:Vacca_apj92} (5) \citet{b:Meurer_apjs06}, (6) \citet{b:Johnson_aj00}, (7) \citet{b:Marlowe_apj95}, (8) \citet{b:Oey_apj07}, (9) \citet{b:Sakai_apj04}}
\tablenotetext{c}{Hubble flow distances for which $H_0=70 \rm km s^{-1} Mpc^{-1}$}
\end{deluxetable}}

{ 
\begin{deluxetable}{l c c c c }
  \tablewidth{0pt}
  \tabletypesize{\footnotesize}
  \tablecaption{Table of Observations \label{t:obs}}
  \tablehead{\colhead{Galaxy} &
  			\colhead{band} & 
             \colhead{$\lambda$} &
             \colhead{Exp time} &             
             \colhead{$3\sigma $} \\
             \colhead{Name} &
             \colhead{} & 
             \colhead{\AA} &
             \colhead{[s]} &
             \colhead{[erg/s/cm$^{2}$/\arcsec]} }
\startdata
\\
\multicolumn{5}{c}{$7-11\ \rm July\ 2009$}\\ \\
\hline
NGC 178 &\Ha  &6594 & $3\times1200$ & $8.46\times 10^{-17} $\\ 
        &\sii &6749 & $4\times1200$ & $1.47\times 10^{-16}$ \\
        &\siii&9113 & $5\times1200$ & $7.36 \times 10^{-17} $\\
        &     &     & $1\times1800$ & \\
NGC 7126&\Ha  &6628 & $3\times1200$ & $9.98\times 10^{-17}$ \\
        &\sii &6783 & $4\times1200$ & $8.25\times 10^{-17}$ \\
        &\siii&9159 & $4\times1200$ & $4.84\times 10^{-17}$\\
\hline \\
\multicolumn{5}{c}{$12-14\ \rm Feb\ 2010$}\\ \\
\hline
NGC 1705&\Ha  &6572$^a$ & $1\times1200$& $9.97\times 10^{-17}$ \\
        &\sii &6730 & $2\times1200$ & $4.10\times 10^{-17}$ \\
        &\siii&9087 & $3\times1200$ & $6.22\times 10^{-17}$ \\
He 2-10 &\Ha  &6582 & $1\times1200$& $1.33\times 10^{-17}$ \\
        &\sii &6735 & $3\times1200$ & $7.01\times 10^{-17}$ \\
        &\siii&9095 & $3\times1200$ & $5.82\times 10^{-17}$ \\
NGC 3125&\Ha  &6587 & $1\times1200$& $1.42\times 10^{-16}$ \\
        &\sii &6740 & $3\times1200$ & $5.02\times 10^{-17}$ \\
        &\siii&9102 & $3\times1200$ & $3.59\times 10^{-17}$ \\
NGC 1482&\Ha  &6602 & $1\times1200$& $1.32\times 10^{-16}$ \\
        &\sii &6759 & $1\times1200$ & $1.20\times 10^{-16}$\\ 
        &\siii&9121 & $2\times1200$ & $5.36\times 10^{-17}$ 
\enddata
\tablenotetext{a}{The appropriate wavelength for \Ha\ in NGC 1705 should be $\lambda6576$. However, our observation set up had the central $\lambda$ = 6572\AA }
\end{deluxetable}}

\section{Results}\label{s:results}

In this section we present and discuss the ionization parameter maps of the galaxies individually. In section \S \ref{s:discuss}, we discuss the results from the sample as a whole.

\subsection{Galaxies with Extended emission}\label{s:detections}

\subsubsection{NGC 5253}

NGC~5253 is a nearby dwarf galaxy currently undergoing an extreme burst of star formation. Using \siii/\sii\ ratio maps, \citet{b:Zastrow_apj11} discovered an optically thin ionization cone in this amorphous elliptical galaxy \citep[Figure 2 of][]{b:Zastrow_apj11}. The ionization cone is narrow with an opening angle $\sim40\degr$ and is detected out to $\sim 800$ pc from
the starburst. The measured opening angle implies a solid angle for
the cone that spans $\sim 3\%$ of 4$\pi$ steradians on the sky,
assuming axisymmetry about the cone axis  \citep{b:Zastrow_apj11}.  If many starbursts
have similar geometry, this suggests that an orientation bias may limit
our ability to detect escaping LyC in starburst and high-redshift galaxies.

\subsubsection{NGC 3125}

\begin{figure}[h]
\centering
\includegraphics[width=3.25in]{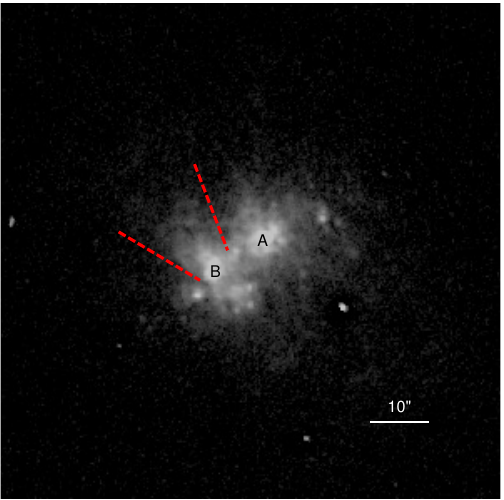}
\includegraphics[width=3.25in]{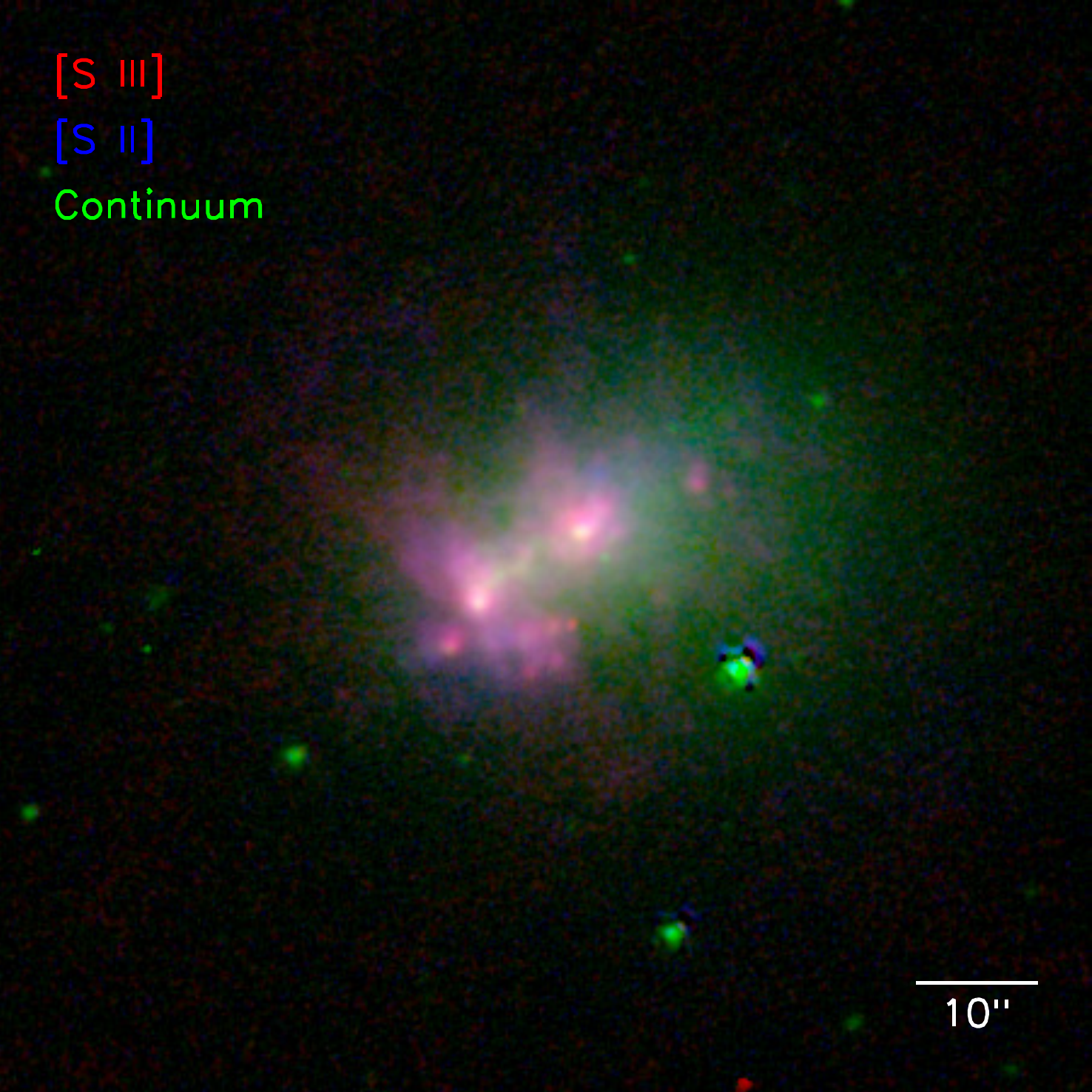}
\caption{Emission-line images of NGC 3125.  Top: \Ha. The red lines illustrate the bounds used to measure the opening angle of the cone.  Bottom: Three-color composite with rest frame \siii$\lambda9069$, \sii$\lambda6716$, and continuum at $\lambda6680$ in red, blue, and green, respectively. At 11.5 Mpc, 10\arcsec = 560 pc. In this figure, N is up and E is to the left. \label{f:img_n3125}}
\end{figure}

\begin{figure}[h]
\subfigure[\siii/\sii]{
\includegraphics[width=3in]{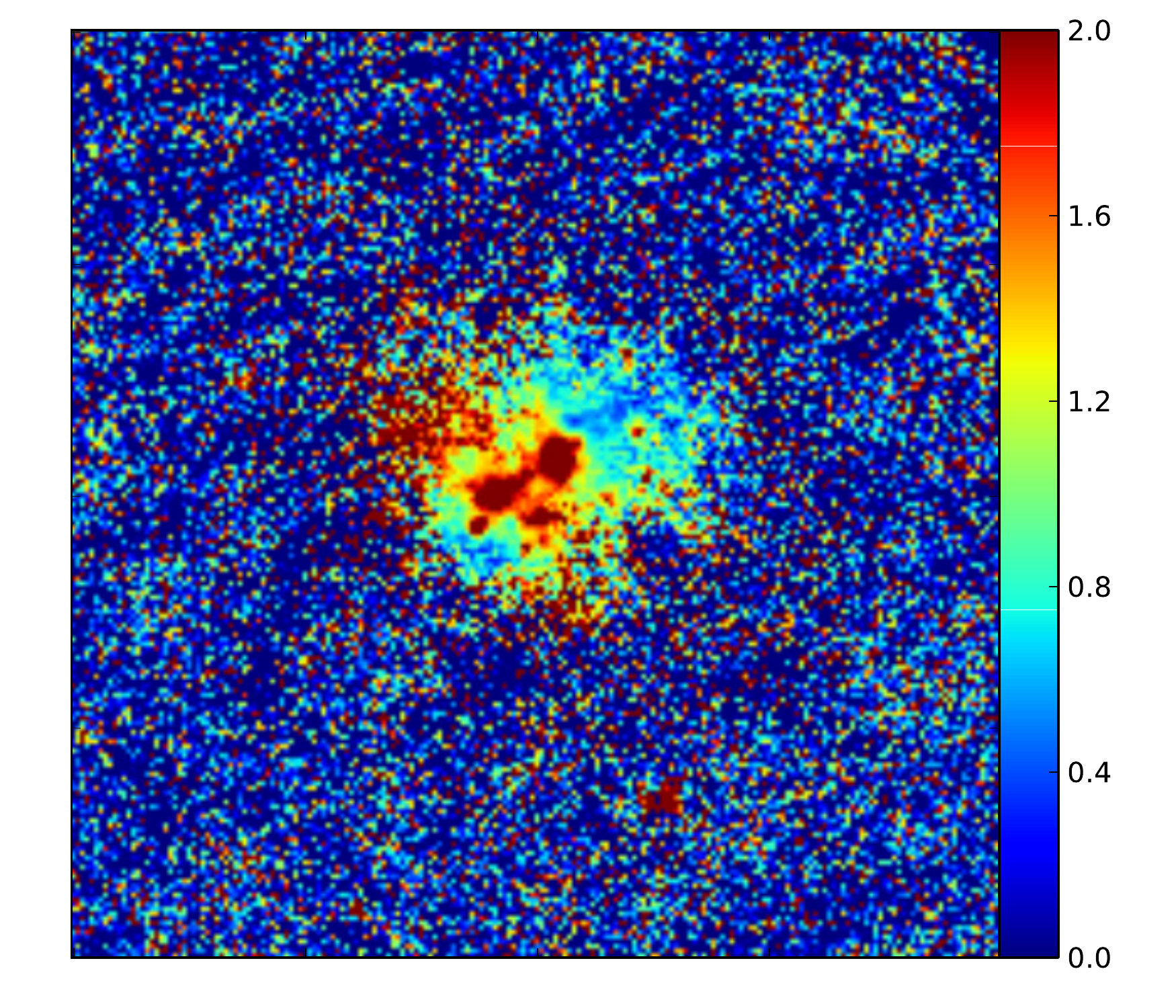}
\label{f:s3ds2_3125}}\\
\subfigure[\siii/\Ha]{
\includegraphics[width=3in]{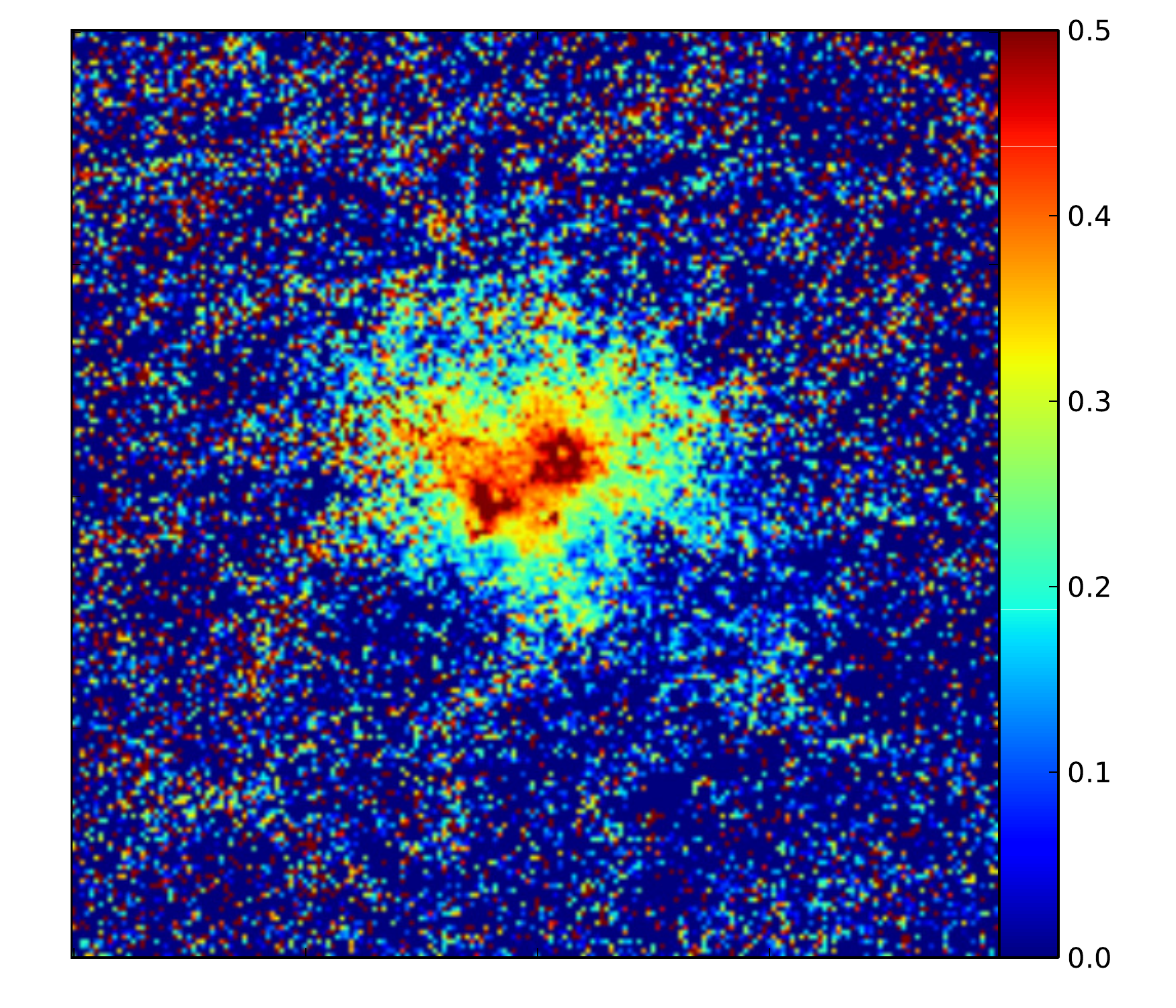}
\label{f:s3dha_3125}}\\
\subfigure[\sii/\Ha]{
\includegraphics[width=3in]{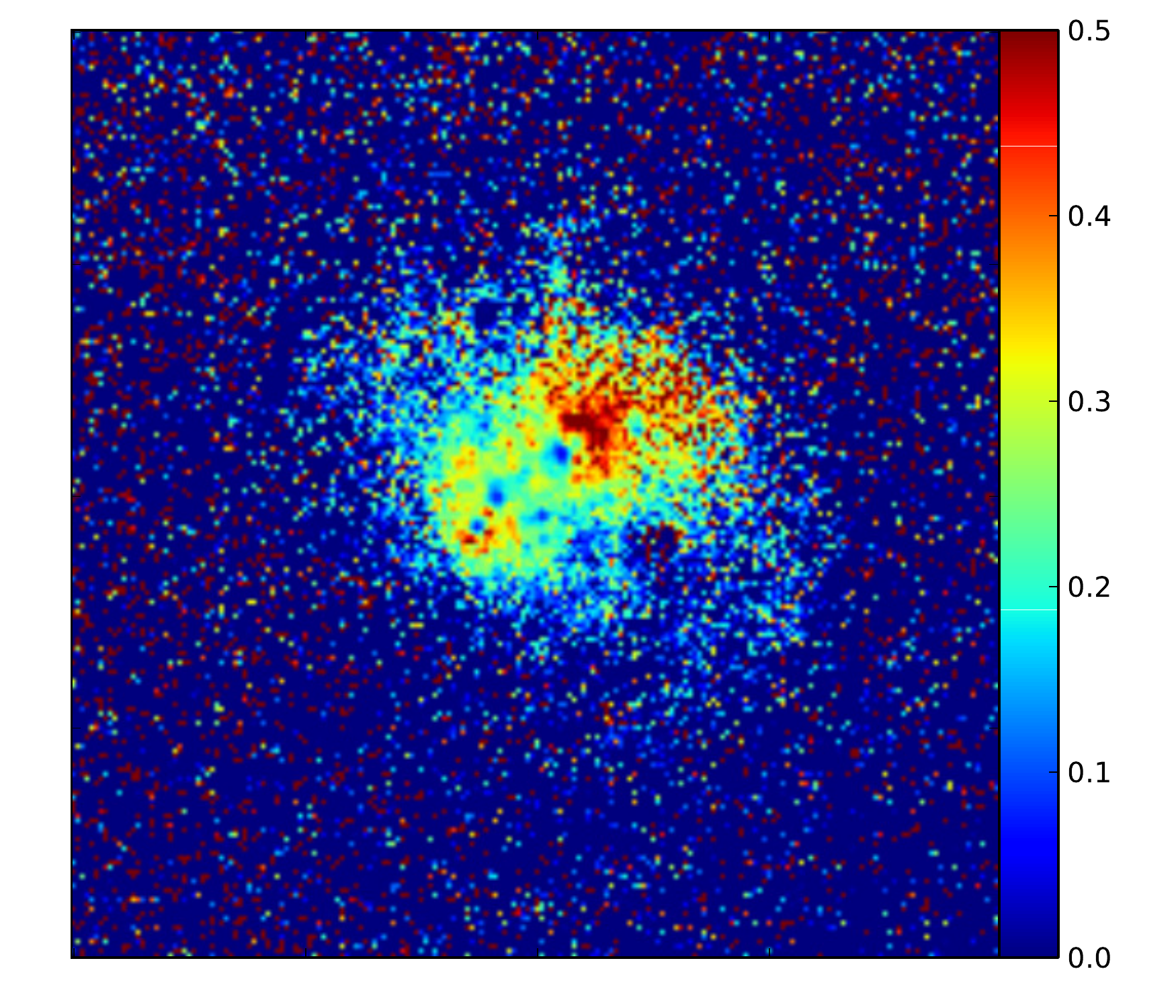}
\label{f:s2dha_3125}}
\caption{Emission-line ratio maps of for NGC 3125. These images have been binned 3x3, resulting in 0.6\arcsec\ pixel sizes. The maps are 50\arcmin\ on a side with an orientation such that North is up and East is to the left.}
\label{f:ratios_3125}
\end{figure}

NGC 3125 is a dwarf galaxy that has a bursty star formation history
with roughly three recent epochs of star formation
\citep{b:Vanzi_aap11}.  This galaxy has two prominent star forming knots,
knot A to the northwest and knot B to the southeast
\citep{b:Vacca_apj92}; both contain massive star clusters and are the
main hosts for the Wolf-Rayet population in the galaxy.  The ionized
gas in NGC~3125 is concentrated around the star-forming knots towards
the eastern side of the galaxy relative to the overall stellar
distribution (Figure \ref{f:img_n3125}).  Figure \ref{f:ratios_3125}
shows the emission-line ratio maps for NGC~3125.  

We discover a bipolar ionization cone in the emission-line gas extending northeast and southwest of knot B.  To the northeast this ionization cone has an opening angle of $40\pm5\degr$. We obtain this measurement using the lines parallel to the edges of the cone, which are shown by the red dotted lines on the \Ha\ image in Figure \ref{f:img_n3125}.  As can be seen in the \siii/\sii\ ratio map (Figure \ref{f:s3ds2_3125}), the ionization cone to the northeast of knot B exhibits high \siii/\sii\ throughout.  The lack of transition to low values of \siii/\sii\ at the edge implies that the ionization cone is optically thin.  This conclusion is supported by the low \sii/\Ha\ observed throughout this portion of the cone (Figure \ref{f:s2dha_3125}), which is consistent with expectations for optically thin gas \citep{b:Pellegrini_apj12}. To the southwest, the line ratios are less conclusive.  While they transition to low \siii/\sii\ as one moves away from knot B, there seems to be another transition to high \siii/\sii\ near the outer edge.  However, the line ratios are less certain in this region.  

\begin{figure}[ht]
\centering
\includegraphics[width=3.45in]{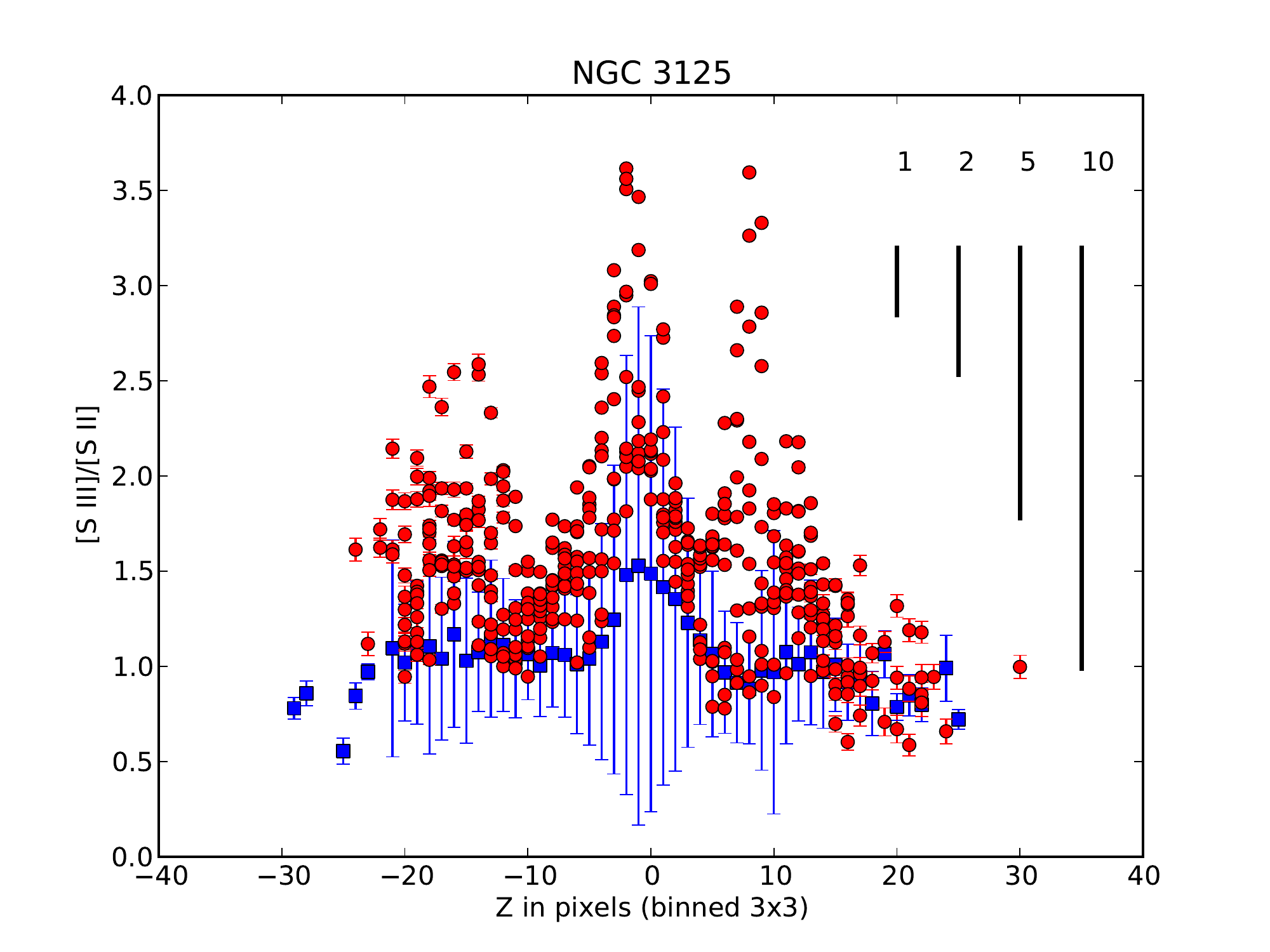}
\includegraphics[width=3.45in]{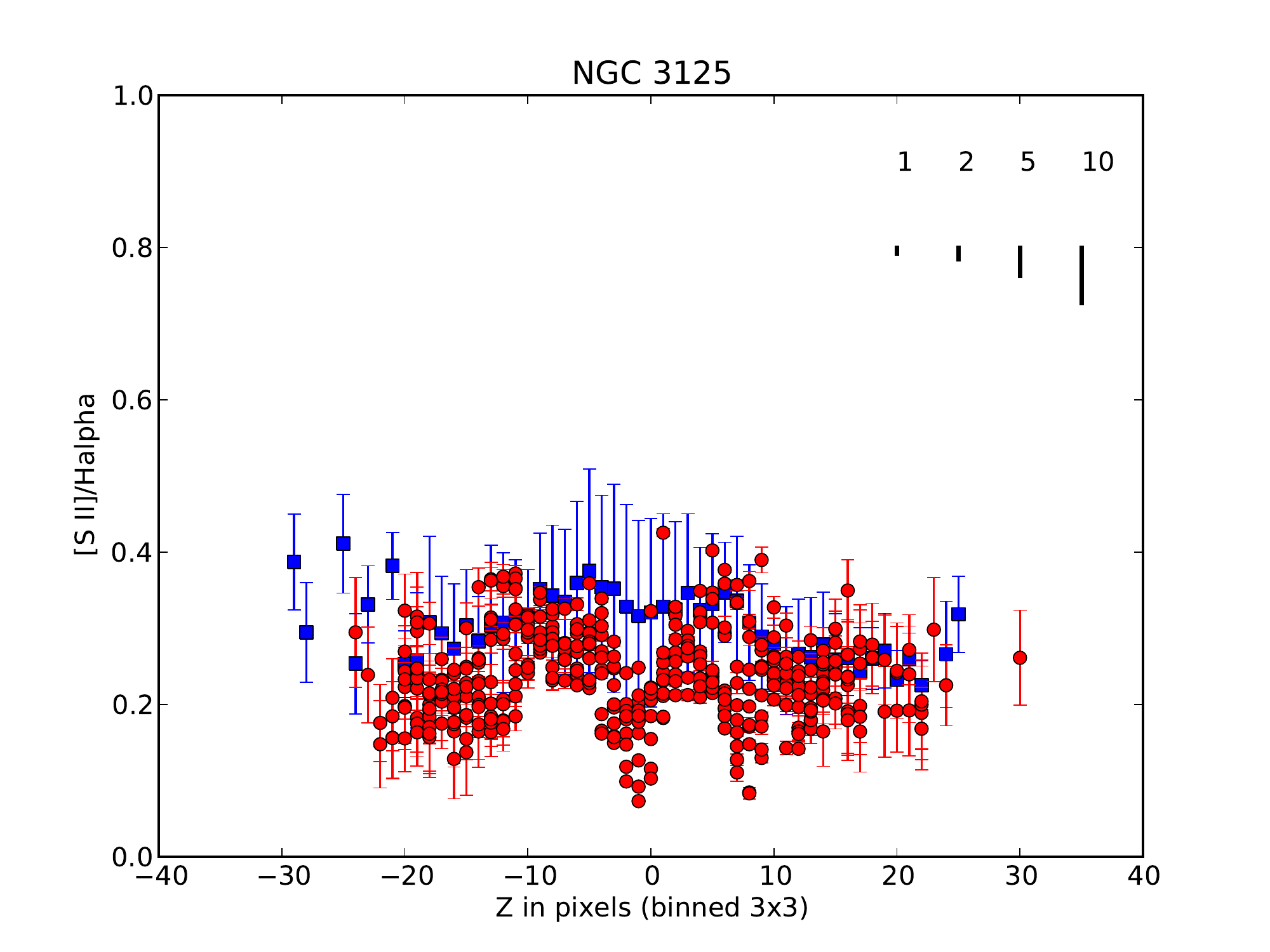}
\includegraphics[width=2.5in]{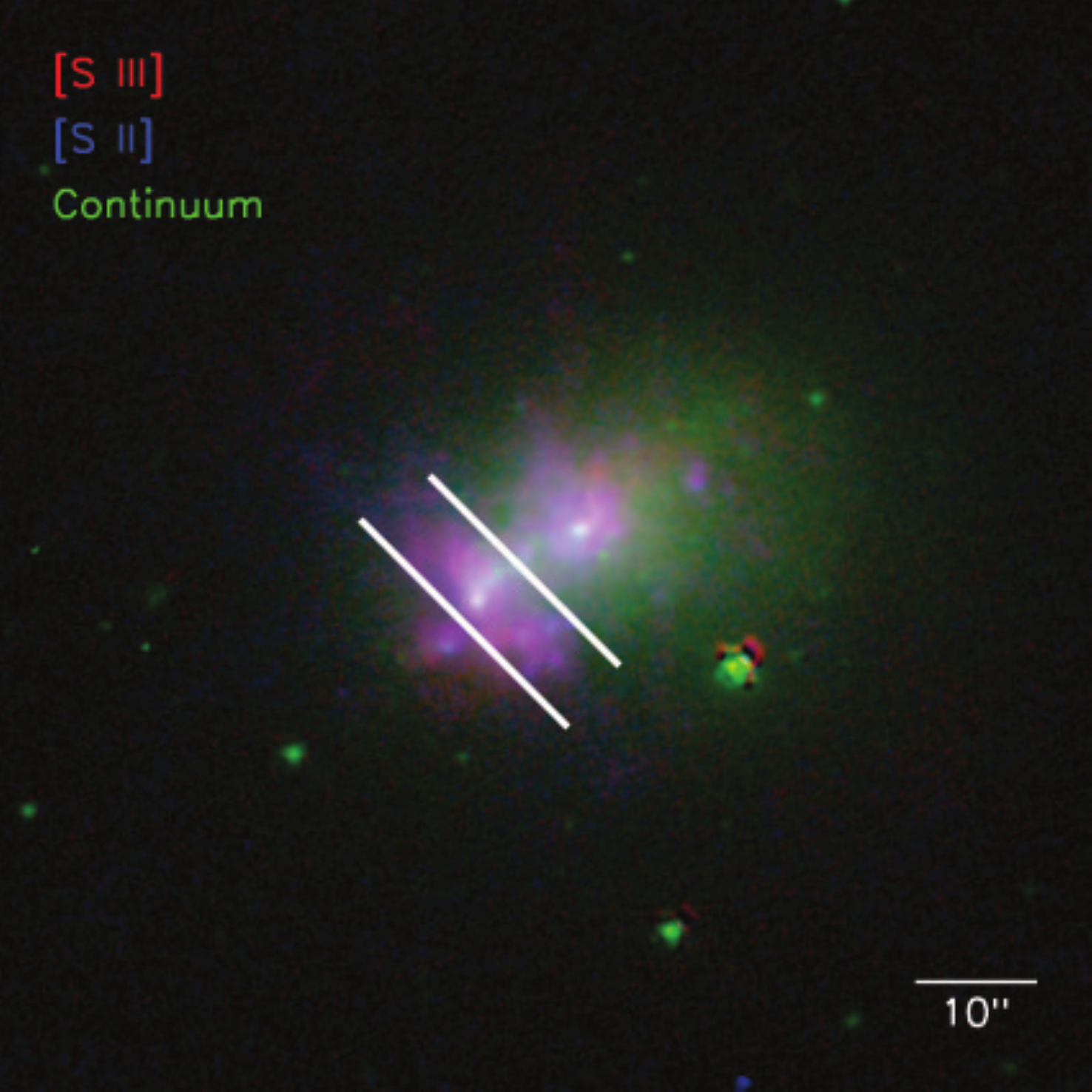}
\caption{Line ratios as a function of projected distance, $z$, above and below the major axis of NGC 3125 for the ionization cone (red points) and the mean line ratio at that $z$ from rest of the galaxy (blue points). \textit{Top}: \siii/\sii. The lines shown on the upper right indicate how the line ratios would change due to reddening of A$_V$ = 1, 2, 5, and 10. All points represent pixels that have a minimum 2$\sigma$ detection in all three emission-line bands. Negative values are towards the northeast, positive values toward the southwest.  \textit{Middle}: Same as the top plot but for \sii/\Ha. \textit{Bottom}: 3-color emission-line image from Figure \ref{f:img_n3125} with white lines delineating the region from which the red points are selected. This Figure demonstrates that there is a clear excess of \siii/\sii\ and a deficiency of \sii/\Ha\ in the NE side of the cone.\label{f:n3125_quant}}
\end{figure}

In Figure \ref{f:n3125_quant}, we quantitatively compare the line ratios in the ionization cone to those of the rest of the galaxy.   Figure \ref{f:n3125_quant} shows the line ratios from the binned ratio maps (Figure \ref{f:ratios_3125}) as a function of $z$, the projected distance above and below the major axis of the galaxy.  The red points represent the line ratios from the ionization cone, selected from the region between the white lines shown in the bottom panel of Figure \ref{f:n3125_quant}.  The blue points are the mean line ratios from the rest of the galaxy at that $z$. The error bars on the blue points represent either the standard deviation of the line ratios at that $z$ or the error on the mean, whichever is more.  All points originate from pixels that have a minimum 2$\sigma$ detection in all three emission-line bands in the 3$\times$3 binned images.  From the top and middle plots in Figure \ref{f:n3125_quant}, we see a clear excess in \siii/\sii\ and a deficiency of \sii/\Ha\ in the NE side of the cone. This behavior is consistent with expectations for optically thin gas. 

The trends in the line ratios discussed above are robust, even after considering the effects of internal reddening on the ratio maps.  The lines shown on the upper right of Figure \ref{f:n3125_quant} indicate how the line ratios would change with different levels of reddening.  In order to explain the observed \siii/\sii\ excess, the reddening would have to have $A_V \sim 5$.  We note that the measured reddening toward knots A and B are $E(B-V) = 0.24$\ and 0.21, respectively \citep{b:Hadfield_mnras06}, which correspond roughly to $A_V \sim 1$. Based on polarimetric observations most of the dust is located in the plane of the galaxy \citep{b:Alton_mnras94}. Thus it is reasonable to expect that the extinction in the cone is much less than that toward knots A and B. Furthermore, while strong reddening would enhance the \siii/\sii\ ratio, correcting for it would drive the \sii/\Ha\ ratio even lower, which strengthens the argument for optically thin gas.

\citet{b:Marlowe_apj95} study the kinematics of the \Ha\ gas in NGC~3125 using echelle spectra.  They detect a Doppler ellipse along the ionization cone to the northeast, showing that it is a bubble expanding with $v = 50$\ km/s.  This detected bubble is consistent with the idea that massive star feedback clears low-density passages in the ISM through which ionizing radiation may escape. To add further support to this picture, low-resolution X-ray observations show that the X-ray distribution is elongated along the minor axis of the galaxy \citep{b:Fabbiano_apjs92}, suggestive of feedback related activity.  Furthermore, the ionization cone occurs at the edge of the galaxy, which may make it easier for ionizing radiation to escape \citep{b:Gnedin_apj08,b:Paardekooper_aap11}. The kinematics towards the southwest are less clear, and \citet{b:Marlowe_apj95} do not detect a Doppler ellipse in this region. 

We also see ionized gas extending to the northeast of knot A. Figure \ref{f:s3ds2_3125} shows that while the gas near the cluster has high ionization parameter, it clearly transitions to a low ionization parameter as the distance from the cluster increases.  The transition indicates that the galaxy is optically thick in that direction.  The line ratio comparison in Figure \ref{f:n3125_quant} supports this analysis.

As discussed in \S \ref{s:data}, our \Ha\ observations are missing a
significant amount of flux due to a combination of wavelength
calibration uncertainties, velocity offsets, and shallow integration.  For
this reason, we use the \Ha\ observations of
\citet{b:GildePaz_apjs03}, to measure the \Ha\ flux in the ionization
cone.  After correcting for an expected 10\% contribution to the flux
from the nearby \nii\ lines, we obtain 
$F(\Ha)= 2.0\ \pm 0.6\ \times 10^{-13}$\ erg s$^{-1}$ cm$^{-2}$, which corresponds to a rate of ionizing photons of log(\Qo)=51.02.  This means that the equivalent of at least 100 ($\pm 30$) O7 stars ionize the cone, assuming \Qo = $10^{49}$ for an O7 star at the metallicity of NGC~3125 \citep{b:Smith_mnras02}.    Knot B has the equivalent of $\sim 1300$ O7 stars based on \Ha\ flux \citep[][adjusted to the same \Qo\ per O7 star]{b:Hadfield_mnras06}. Based on these numbers,  roughly 8\% of the ionizing radiation generated by the cluster needs to escape into the NE cone.  

The ionization cone in NGC~3125 is narrow and appears somewhat collimated, with an opening angle of $40\pm5\degr$.  This angle corresponds to a solid angle of $3\pm1$\% of 4$\pi$ steradians, assuming axisymmetry about the cone axis. We note that this value is not inconsistent with the 8\% escape fraction estimated from the \Ha\ flux, above. Since knot B is smaller than the base of the cone, the covering fraction of the cone relative to the cluster is larger than that of the cone on the sky. We note that the 8\% quoted here only refers to the cluster in knot B.  Using the total $F$(\Ha)=3.4$\times 10^{-12}$ erg s$^{-1}$ cm$^{-2}$ \citep{b:Marlowe_apj95}, the cone represents $\sim6\%$ of the ionizing flux of the galaxy.

The presence of an optically thin ionization cone suggests that ionizing radiation is escaping this starburst. At the very least, we see that ionizing radiation is escaping the main body of the galaxy and traveling into the halo. The small implied solid angle, in conjunction with a preferred direction is similar to the geometry for NGC 5253, again supporting the possibility that an orientation bias affects our ability to detect escaping Lyman continuum in starbursts and LBGs. 

\subsubsection{NGC 1705}

NGC 1705 is a nearby starburst galaxy with a stunning starburst-driven galactic wind \citep{b:Meurer_aj92}.  NGC 1705 has a combined stellar and gas mass of $\sim2.7\times10^8$\Msun\ \citep{b:Meurer_aj92}.  The metallicity measurements range from 12+log(O/H) = 8.21 to 8.46 \citep{b:Lee_apj04,b:Storchi-Bergmann_apj94,b:Meurer_aj92}. Like similar dwarf starbursts, NGC 1705 has a bursty star formation history with two recent star forming epochs in the last 15 Myr \citep{b:Annibali_aj03,b:Annibali_aj09}.  One of these bursts formed the well known super-star cluster (SSC) at the center, which has log($M$)$\sim 10^5$ and an age of $\sim 10$\ Myr  \citep{b:Melnick_aap85,b:OConnell_apj94,b:Ho_apj96}.  This SSC generates most of the UV luminosity and is consistent with a population dominated by B stars \citep{b:Heckman_aj97}. That same 10--15 Myr old burst of star formation is thought to have launched the galactic wind \citep{b:Annibali_aj03,b:Annibali_aj09}. This burst was followed by a more recent epoch of star formation, $\sim 3$\ Myr ago that may have been triggered by the wind \citep{b:Annibali_aj03,b:Annibali_aj09}. 

\begin{figure}[h]
\centering
\includegraphics[width=3.45in]{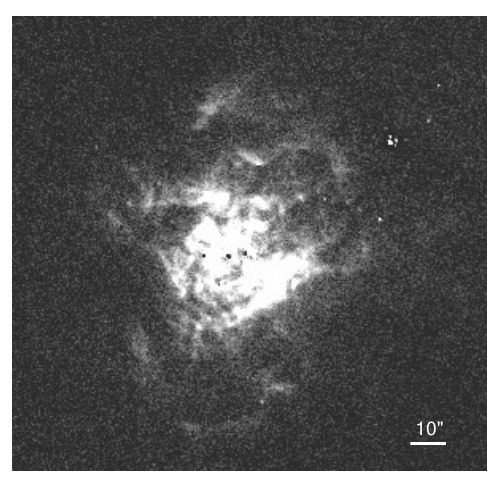}
\includegraphics[width=3.25in]{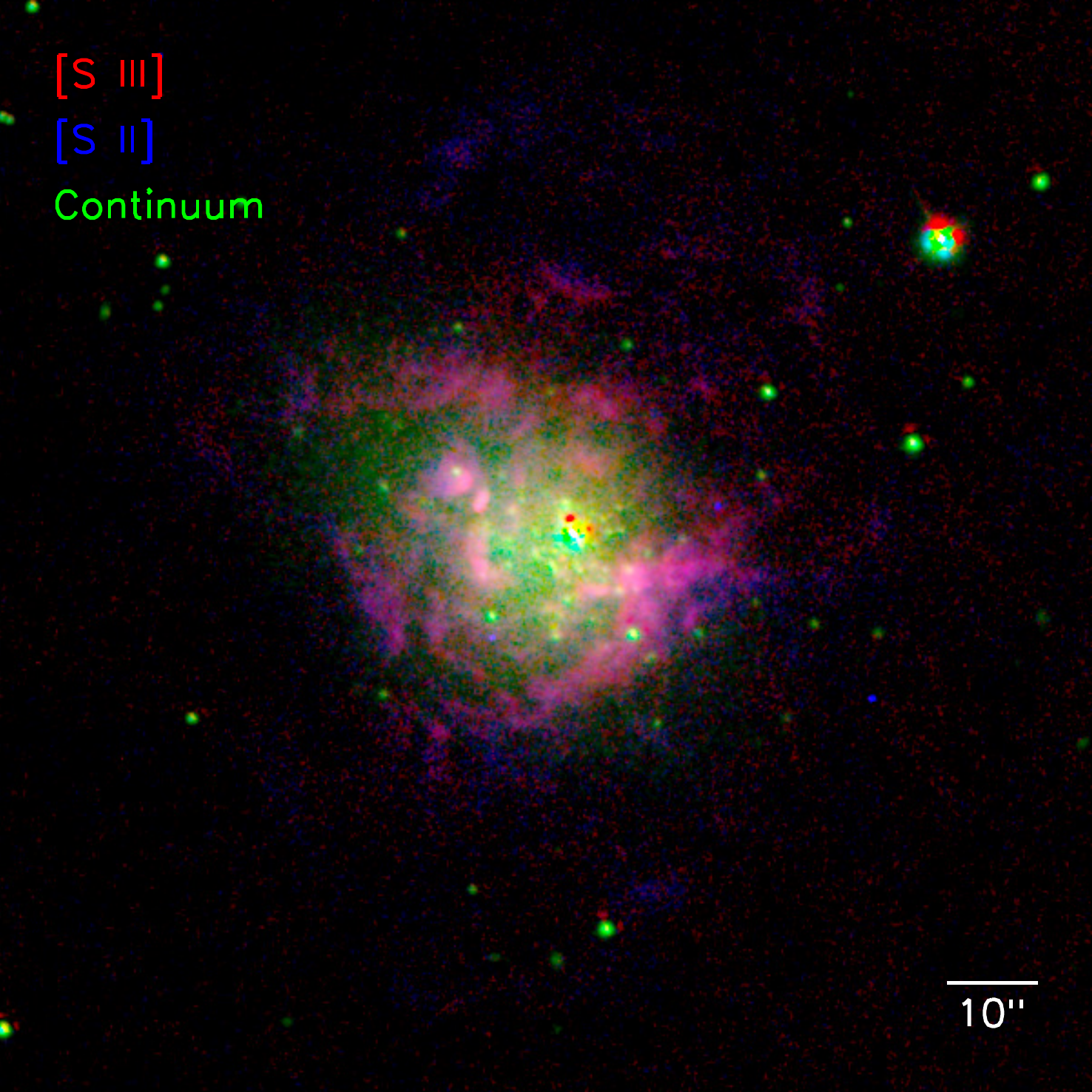}
\caption{Emission-line images of NGC 1705.  Top: \Ha\ Bottom: Three-color composite with rest frame \siii$\lambda9069$, \sii$\lambda6716$, and continuum at $\lambda6680$ in red, blue, and green, respectively. At 5.1 Mpc, 10\arcsec = 250 pc. In this figure, N is up and E is to the left. \label{f:img_n1705}}
\end{figure}

\begin{figure}[h]
\subfigure[\siii/\sii]{
\includegraphics[width=3in]{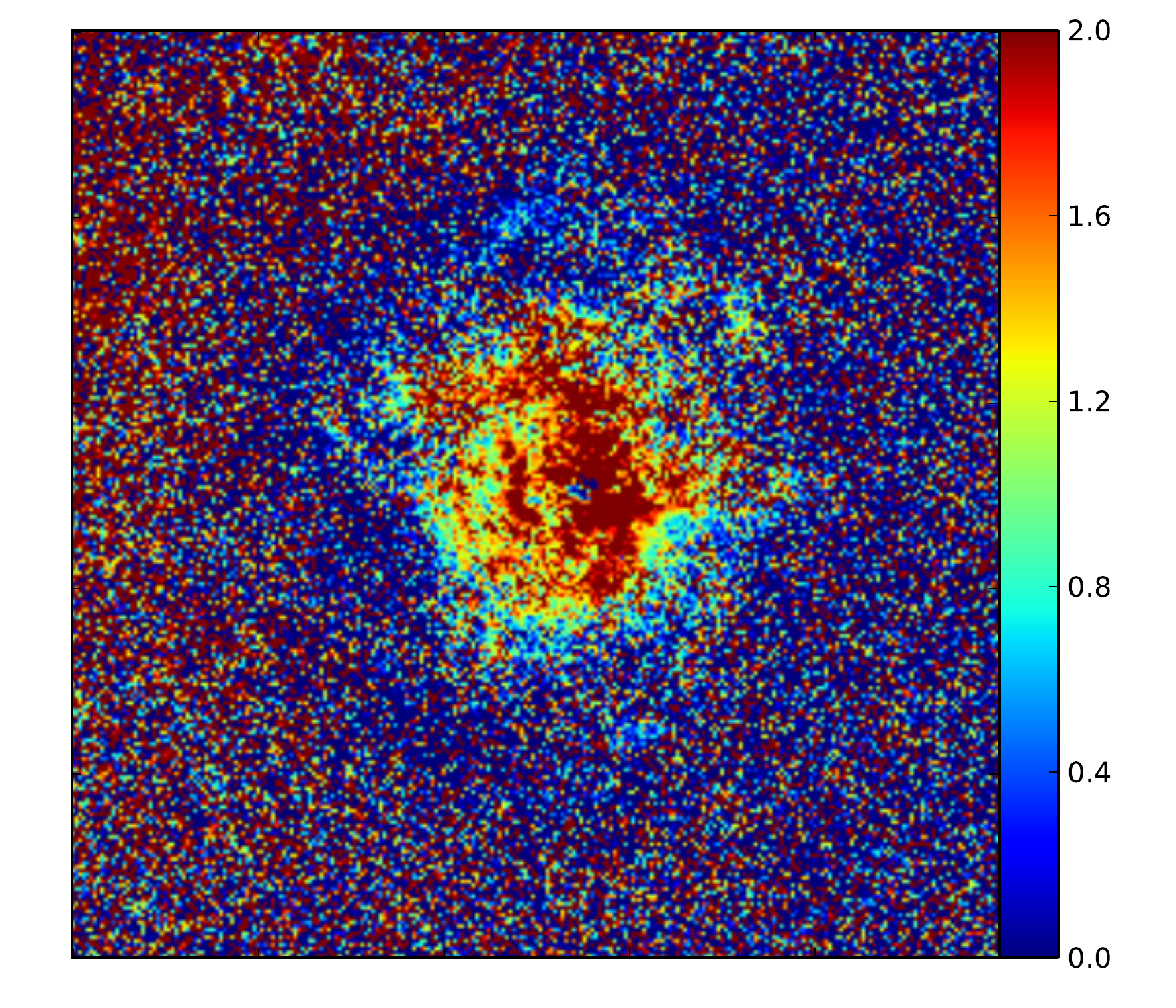}
\label{f:s3ds2_1705}}\\
\subfigure[\siii/\Ha]{
\includegraphics[width=3in]{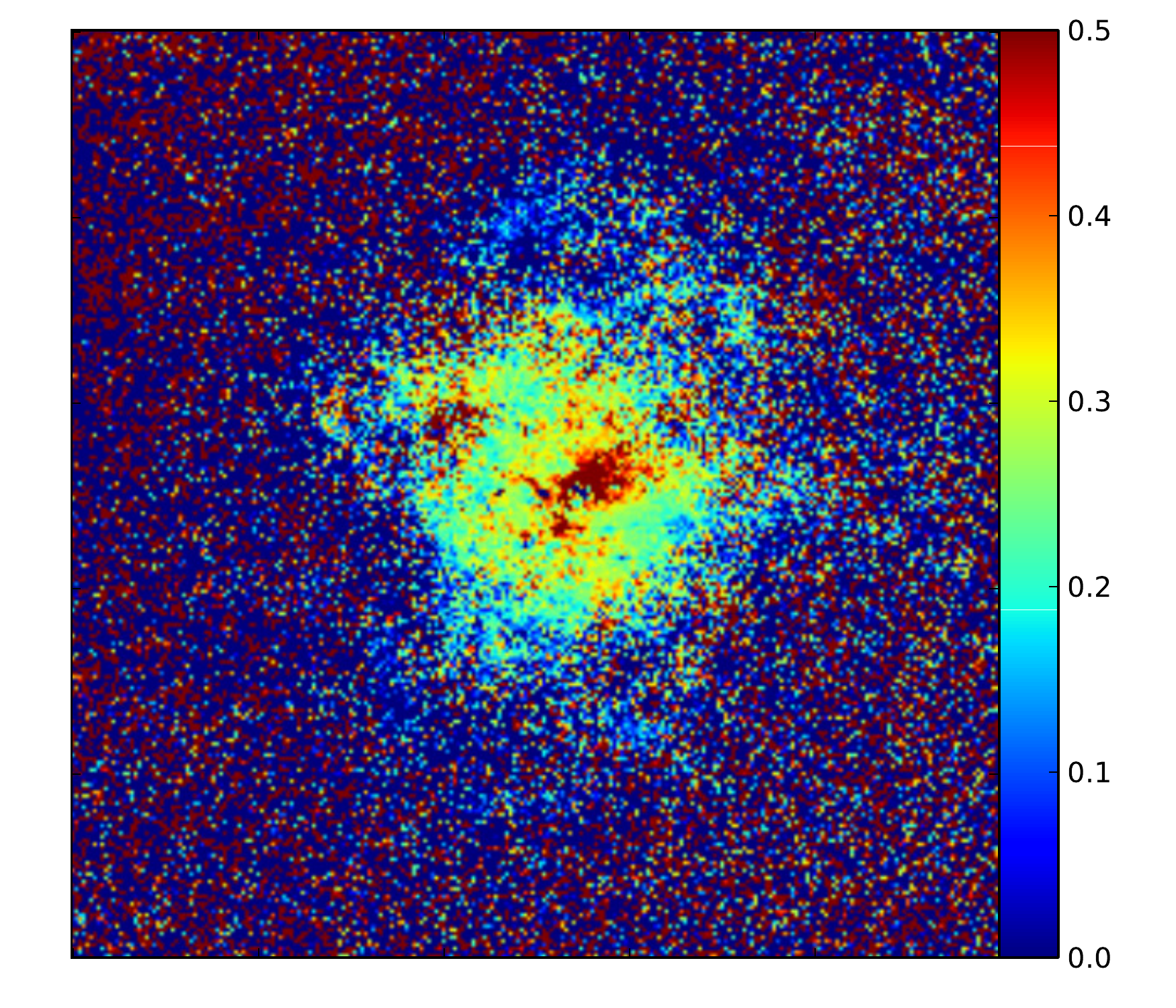}
\label{f:s3dha_1705}}\\
\subfigure[\sii/\Ha]{
\includegraphics[width=3in]{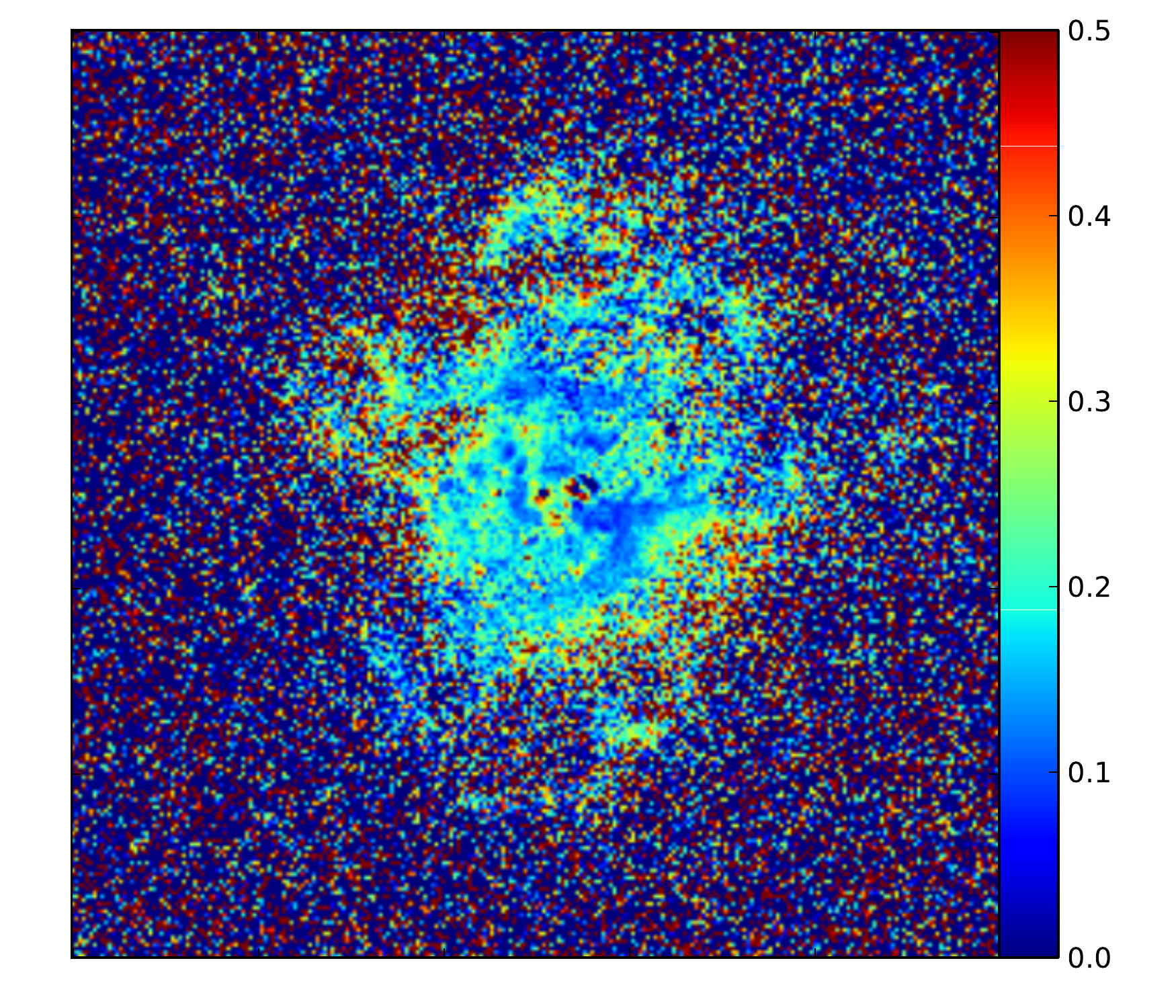}
\label{f:s2dha_1705}}
\caption{Emission-line ratio maps of for NGC 1705. These images have been binned 3x3, resulting in 0.6\arcsec\ pixel sizes. The maps are 62.5\arcmin\  on a side with an orientation such that North is up and East is to the left.}
\label{f:ratios_1705}
\end{figure}

In emission-line gas, NGC~1705 shows filamentary extended substructure as a result of its galactic wind \citep{b:Meurer_aj92}.  Based on \Ha\ observations, the superbubble has an expansion velocity 50--132 km/s \citep{b:Marlowe_apj95,b:Meurer_aj92}. \citet{b:Heckman_aj97} studied the kinematics using UV interstellar absorption lines and found further evidence for an expanding wind. X-ray observations show that the wind has similar size and morphology to the optically detected gas \citep{b:Strickland_apjs04}. The star-forming nucleus resides on the axis of symmetry for the superbubble, which further supports that the wind is driven by star formation \citep{b:Meurer_aj92}.  Shocks contribute only a small amount to the ionization in the wind, as most of the emission-line gas is photo-ionized \citep{b:Marlowe_apj95,b:Veilleux_aj03}. Our emission-line images and ratio maps for NGC~1705 are shown in Figures \ref{f:img_n1705}, and \ref{f:ratios_1705}.  The ionized gas associated with the wind is clearly detected in both \siii\ and \sii.  

The spatial changes in the \siii/\sii\ ratio map suggest that the wind in NGC~1705 is optically thick.  Close to the starburst, the
\siii/\sii\ ratio is high, as expected from photoionized gas near the ionization source. However, as one looks further into the wind, there is a clear transition towards lower \siii/\sii\ ratios, and at the far edge of the shells, the line ratios indicate the \sii\ emission is dominant.  This is further supported by the \sii/\Ha\ and \siii/\Ha\ ratio maps in Figures \ref{f:s2dha_1705} and \ref{f:s3dha_1705}. In these maps, the \sii/\Ha\ ratio increases at the edges of the filament, and, as expected based on Figure \ref{f:s3ds2_1705}, the \siii/\Ha\ ratio decreases. While this behavior is consistent with optically thick gas, we note that if the radiation field is dominated by hot stars with softer ionizing spectra, as in the case of NGC~1705, the gas may simultaneously exhibit a low-ionization transition zone and be optically thin \citep{b:Pellegrini_apj12}.  

In addition to evaluating the optical depth in the plane of the sky, the ratio maps can be used to evaluate the optical depth along the line of sight.  For gas with abundances similar to that of the LMC, \sii/\Ha\ $< 0.05$ is a diagnostic for optically thin photoionized gas along the line of sight.  Of particular interest is the region around NGC~1705-1, the super star cluster.  \citet{b:Heckman_aj97} estimated a very low absorbing \HI\ column, $N_\HI = 1.5 \times 10^{20} \rm cm^{-2}$, toward NGC~1705-1.  This column is an order of magnitude smaller than the column measured from 21 cm observations, which suggests that the cluster is either sitting in front of most of the \HI\ or that we are observing the SSC through a hole in the ISM \citep{b:Heckman_aj97}.  Either scenario increases the likelihood for escaping Lyman continuum. However, Figure \ref{f:s2dha_1705} shows that \sii/\Ha\ $> 0.05$ over the entire nuclear region, which indicates that the nuclear region is optically thick in the line of sight.  The low \HI\ column combined with optical \sii/\Ha\ implies that even though the neutral gas may have large variations in column density, the majority of sight lines are optically thick.

It is expected that strong starburst feedback, such as that present in NGC~1705, is conducive to escaping Lyman continuum.  However, our observations indicate that the wind in NGC 1705 is optically thick. Both the strength of the radiation field and the amount of gas available to absorb the ionizing radiation will determine whether the galactic wind is optically thick or thin. While there have been two recent epochs of star formation, one 10-15 Myr ago and one $\sim 3$ Myr ago \citep{b:Annibali_aj03,b:Annibali_aj09}, the ionizing population is dominated by B stars \citep{b:Heckman_aj97}. This means that there are relatively fewer ionizing photons being emitted and the radiation field is softer than if there were significant numbers of O stars.  In addition to a reduced radiation field, a `spur' of \HI\ gas is associated with the outflow \citep{b:Meurer_mnras98}.  This spur of neutral gas is likely co-spatial with the optically detected galactic wind based on its kinematics and orientation \citep{b:Meurer_mnras98,b:Elson_mnras13}.  The feature has an \HI\ column density of $\sim10^{20}\rm\ cm^{-2}$, which is more than high enough to be optically thick to Lyman continuum.  Thus, despite the presence of a strong galactic wind, NGC~1705 is likely optically thick due to a combination of population age and gas morphology. 

\subsubsection{He 2-10}

\begin{figure}[h]
\centering
\includegraphics[width=3.35in]{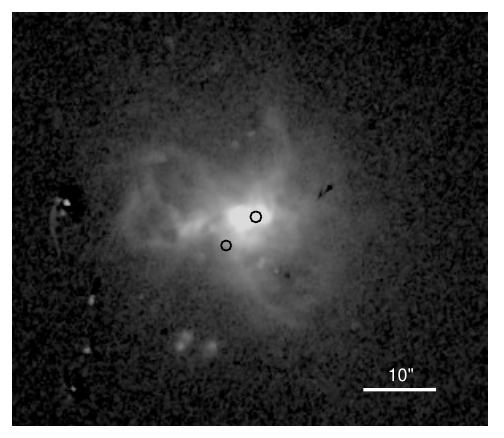}
\includegraphics[width=3.25in]{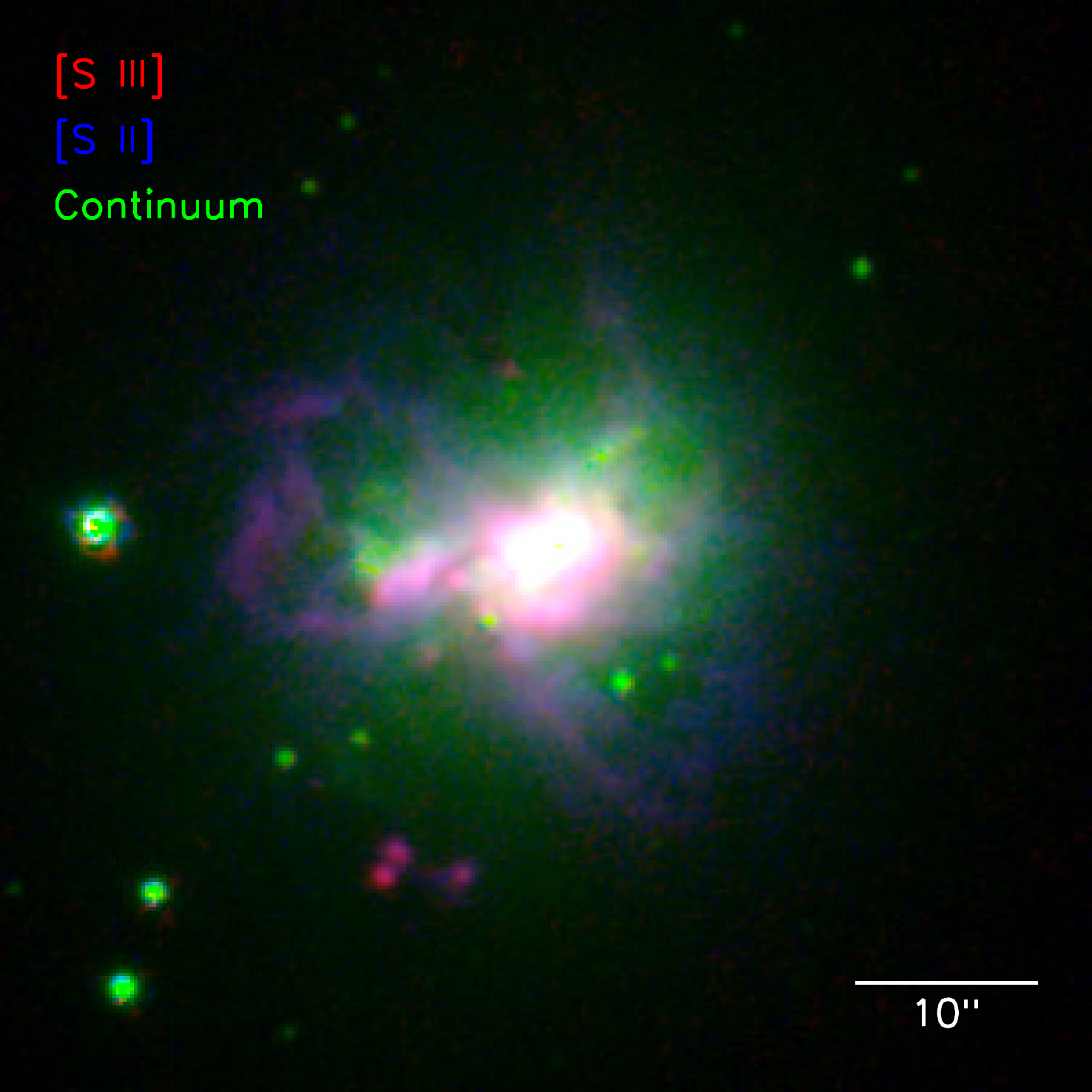}
\caption{Emission-line images of He 2-10. Top: \Ha.  The two circles correspond to regions that have \sii/\Ha\ $<0.05$. Bottom: Three-color composite with rest frame \siii$\lambda9069$, \sii$\lambda6716$, and continuum at $\lambda6680$ in red, blue, and green, respectively. At 9 Mpc, 10\arcsec = 460 pc. In this figure, N is up and E is to the left. \label{f:img_he210}}
\end{figure}

He 2-10 is one of the prototypical HII galaxies \citep{b:Allen_mnras76}. It is 9 Mpc away \citep{b:Vacca_apj92} and has dynamical mass $\sim3 \times 10^9$ \Msun \citep{b:Kobulnicky_aj95}. In emission-line gas there are many notable loops and filaments (Figure \ref{f:img_he210}) whose kinematics suggest expanding bubbles and possible outflows \citep{b:Mendez_aap99}. These bubbles are centered on sites of recent star formation.  He 2-10 is the host of nearly 80 SSCs \citep{b:Johnson_aj00}. The central starburst contains a handful of clusters in ultra-compact \hii\ regions that contain the ionizing equivalent of 700-2600 O7 stars each \citep{b:Johnson_apj03}. In addition to prodigiously forming stars, SFR $\sim$ 0.87 - 2 \Msun/yr \citep[][]{b:Calzetti_apj10,b:Reines_nature11}, this galaxy was recently found to be the host of an AGN \citep[][]{b:Reines_nature11}.

\begin{figure}[h]
\subfigure[\siii/\sii]{
\includegraphics[width=3in]{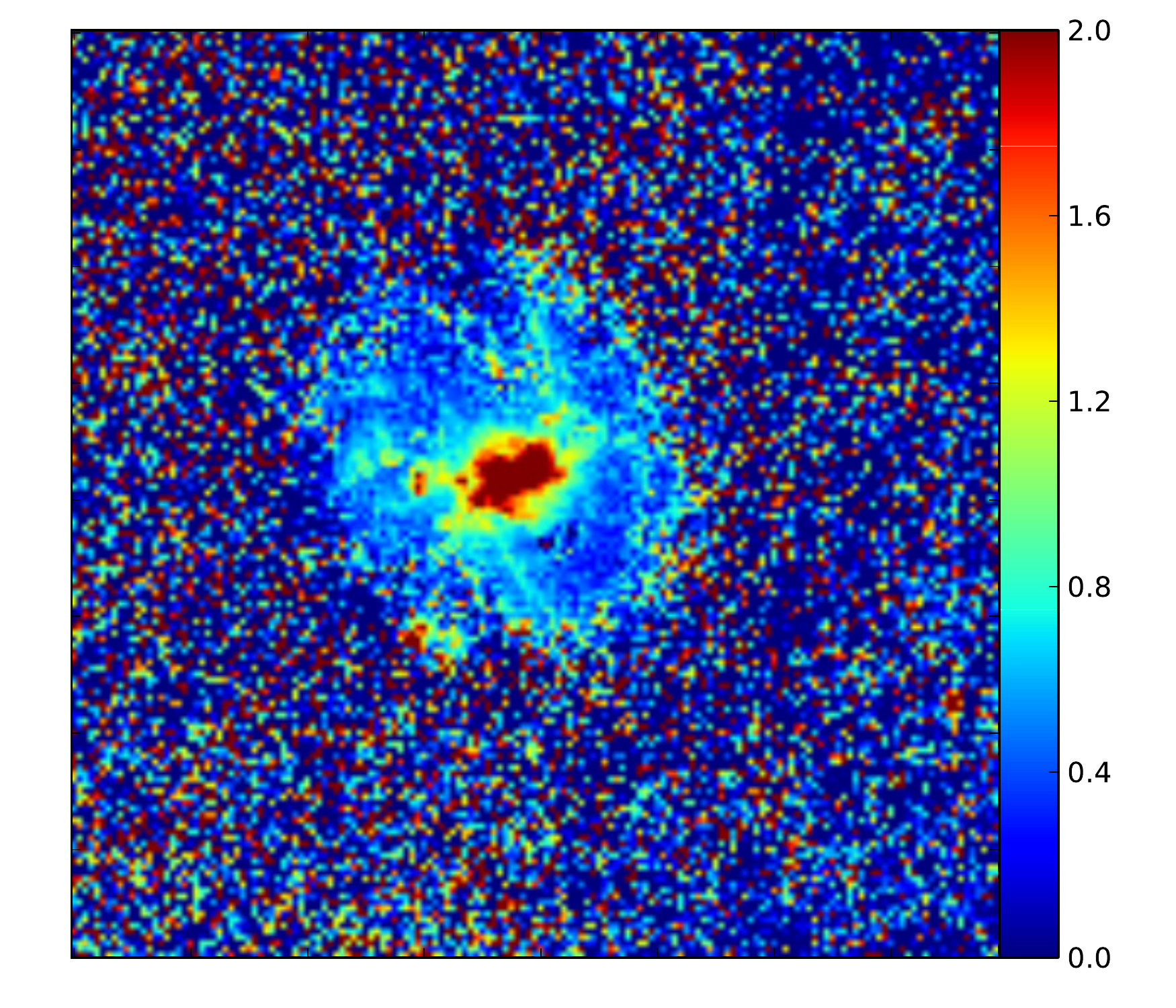}
\label{f:s3ds2_he210}}\\
\subfigure[\siii/\Ha]{
\includegraphics[width=3in]{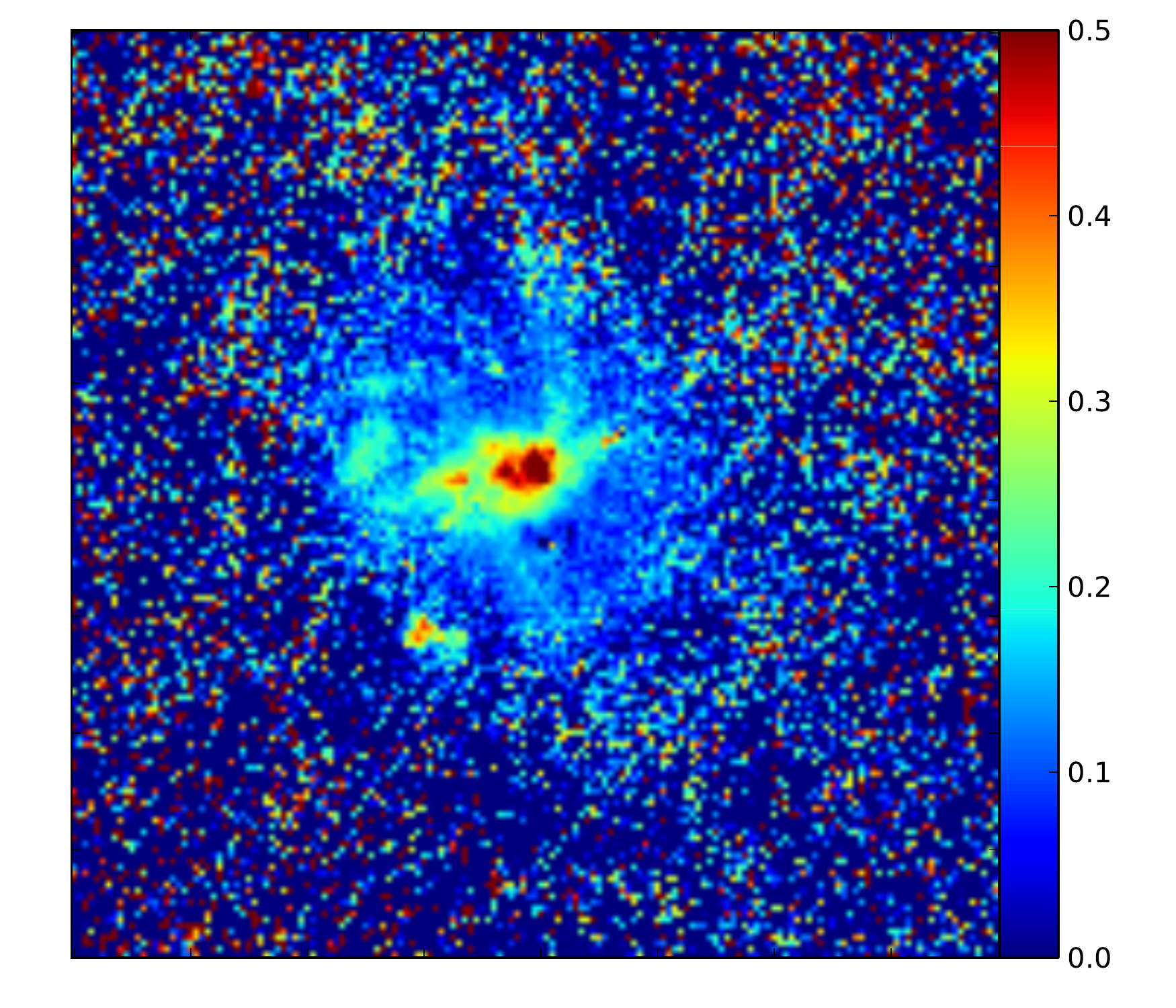}
\label{f:s3dha_he210}}\\
\subfigure[\sii/\Ha]{
\includegraphics[width=3in]{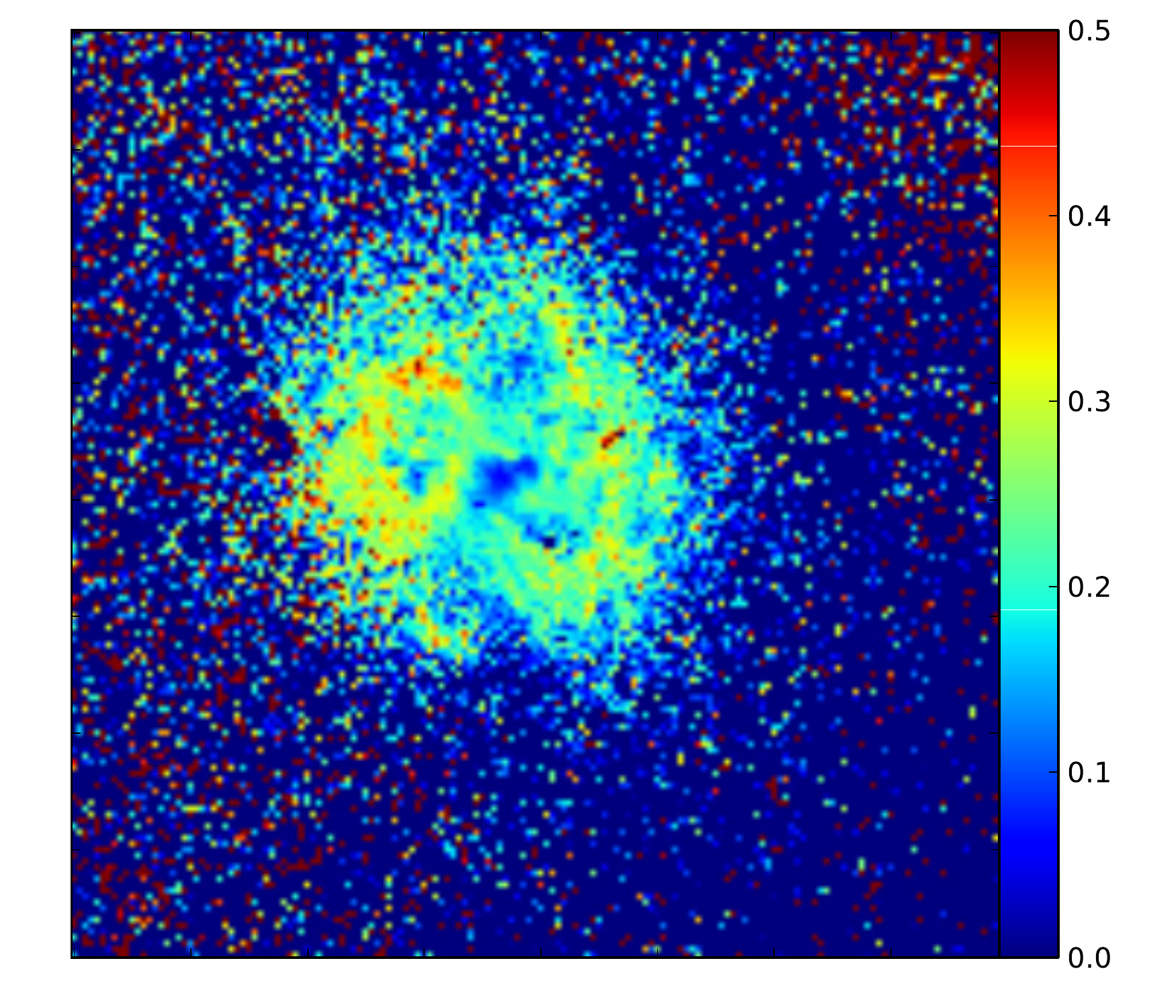}
\label{f:s2dha_he210}}
\caption{Emission-line ratio maps of for He~2-10. These images have been binned 3x3, resulting in 0.6\arcsec\ pixel sizes. The maps are 39.58\arcmin\  on a side with an orientation such that North is up and East is to the left.}
\label{f:ratios_he210}
\end{figure}

In the \siii/\sii\ ratio map (Figure \ref{f:s3ds2_he210}), the central star forming region stands out with high \siii/\sii, as would be expected for a region hosting thousands of O stars.  As the radius from the center increases, \siii/\sii\ drops considerably. There is a local maximum $\sim 8\arcsec$\ to the east around the location of the second SSC.  However, even here the emission-line gas transitions to low \siii/\sii\ toward the edge of the expanding bubble.  The contrast at the edges of the bubbles is seen more clearly in the \sii/\Ha\ ratio map (Figure \ref{f:s2dha_he210}), where a clear enhancement of \sii/\Ha\ is present. In contrast the \siii/\Ha\ ratio (Figure \ref{f:s3dha_he210}) falls off as the distance from the ionizing source increases, as would be expected for an optically thick nebula.

The spatial changes in the line ratio maps are consistent with
expanding superbubbles. High ionization parameter gas is present close to the ionizing source, while the edges of the bubbles have low ionization parameter and possible contribution to \sii\ from shocks. Studies of the \Ha\ kinematics indicate a potential Doppler ellipse in the NE bubble and velocities differences of up to $\pm300$ km/s across the galaxy \citep{b:Mendez_aap99}.  Using UV interstellar lines, \citet{b:Johnson_aj00} determined that the bulk motion of the ISM indicates an outflow with $v\sim360$ km/s.  The velocities reported by both works are in excess of the escape velocity $\sim160\pm30$\ km/s \citep{b:Johnson_aj00}.  However, the closed loops of the outflow seen in the \Ha\ images suggest that these bubbles have not yet broken out of the ISM to form a galactic fountain. Bubble walls severely hinder the passage of ionizing radiation, and can delay the escape of Lyman continuum radiation until $Q_0$ has dropped considerably \citep{b:Dove_apj00,b:Fujita_apj03}.  

He 2-10 is unique in our sample, in that it is the only galaxy that contains a confirmed AGN.  This AGN was discovered by \citet{b:Reines_nature11} as a radio and X-ray point source and is located near the star forming region of knot A. The accretion rate onto the AGN is modest, but may have been greater in the past \citep{b:Reines_nature11}. Thus, the presumed tracers for massive-star feedback, such as the large expanding bubbles, may be contaminated by contributions from the AGN as opposed to being driven by star formation. 

The distribution of gas in the ISM plays a critical role in the passage of ionizing radiation in a galaxy.  The \HI\ observations from \citet{b:Kobulnicky_aj95} show that the neutral ISM extends significantly beyond the optically detected ionized gas. This observation is consistent with the observations of optically thick bubbles. However, if this overall \HI\ envelope has a clumpy distribution, it is still possible to have small low-density passages through which ionizing radiation may escape the starburst.  The \sii/\Ha\ ratio map shows two locations that have \sii/\Ha $< 0.05$ and may be optically thin.  We note these locations with small circles on Figure \ref{f:img_he210}. One of these regions coincides with the location of the AGN, and the other lies 4\arcsec\ to the SE from the first. While our data cannot establish the presence of an optically thin path from the galaxy conclusively, these observations are suggestive that feedback from the AGN or star formation in the region have cleared out a small hole in the ISM. The variable reddening measurements derived from different wavelengths and positions in the galaxy further suggest an inhomogeneous distribution of gas and dust throughout the galaxy \citep[e.g.,][]{b:Sauvage_aap97}.  The presence of small local holes is also supported by the fact that many of the probably young SSCs do not show commensurate \Ha\ emission, which suggests that gas has been evacuated in the regions of these clusters \citep{b:Johnson_aj00}.  If this is the case, it supports the picture that ionizing radiation escapes galaxies through small holes and that an orientation bias affects our ability to detect escaping Lyman continuum.  We note that these line ratios have not been corrected for reddening. However, any reddening correction would drive the \sii/\Ha\ line ratio lower, strengthening the argument for optically thin gas along the line of sight.
 
\subsubsection{NGC 1482}

NGC 1482 is an SA0 \citep{b:deVaucouleurs_rc391} with a well known galactic wind \citep{b:Hameed_aj99}. At a distance of 22.6 Mpc \citep{b:Kennicutt_pasp11}, this starburst is a highly inclined disk with a prominent dust lane \citep{b:deVaucouleurs_rc391}.  The wind was first discovered by \citep{b:Hameed_aj99} in optical emission-lines, with later optical and X-ray observations confirming its nature as a wind \citep[][]{b:Veilleux_apj02,b:Strickland_apjs04}. 

\begin{figure}[h]
\centering
\includegraphics[width=3.45in]{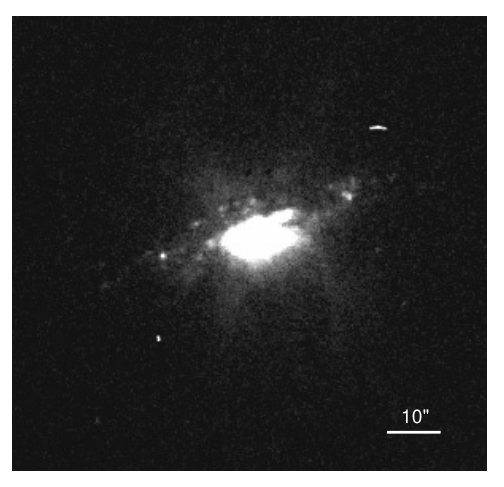}
\includegraphics[width=3.25in]{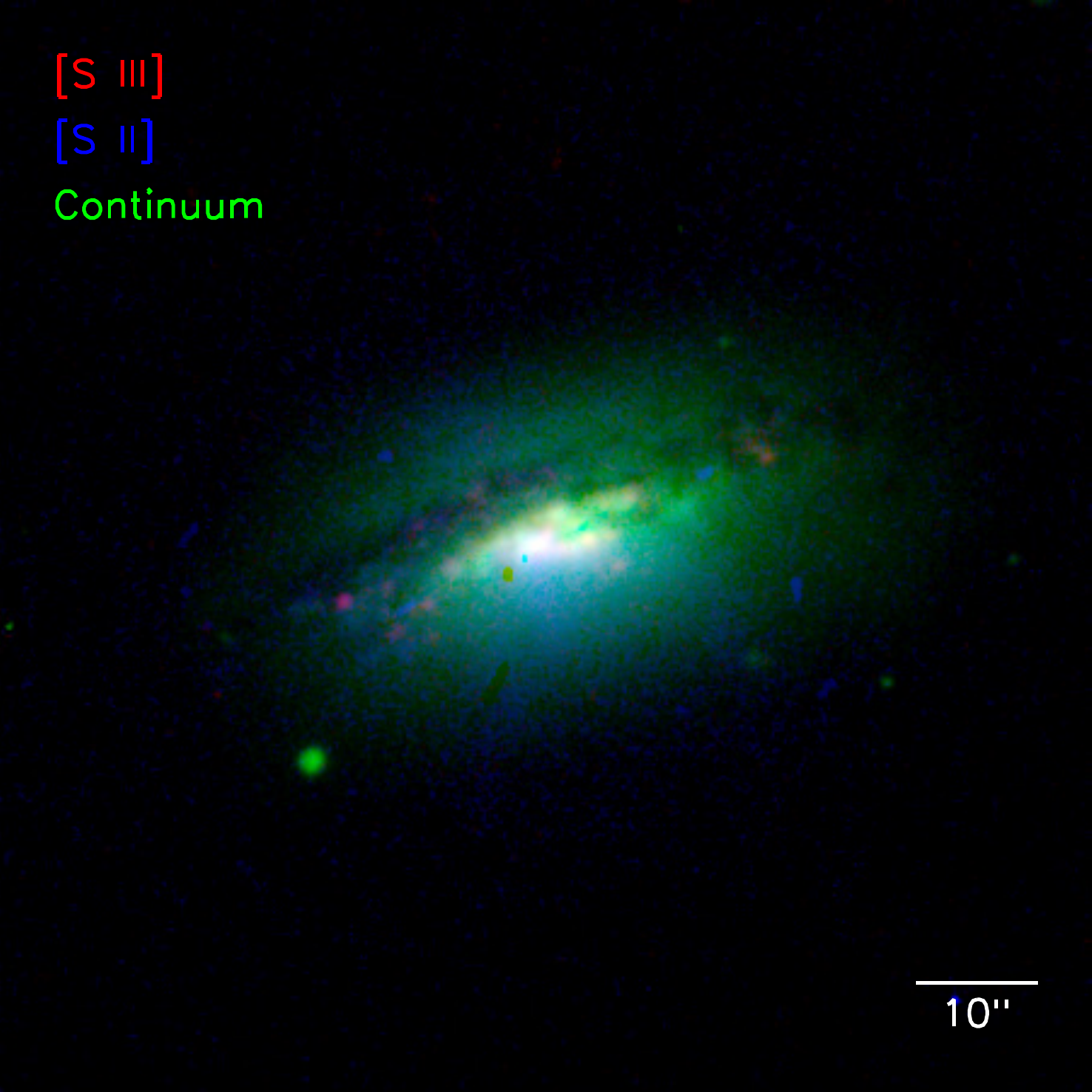}
\caption{Emission-line images of NGC 1482. Top: \Ha\ Bottom: Three-color composite with rest frame \siii$\lambda9069$, \sii$\lambda6716$, and continuum at $\lambda6680$ in red, blue, and green, respectively. At 22.6 Mpc, 10\arcsec = 1.1 kpc. In this figure, N is up and E is to the left. \label{f:img_n1482}}
\end{figure}

\begin{figure}[h]
\subfigure[\siii/\sii]{
\includegraphics[width=3in]{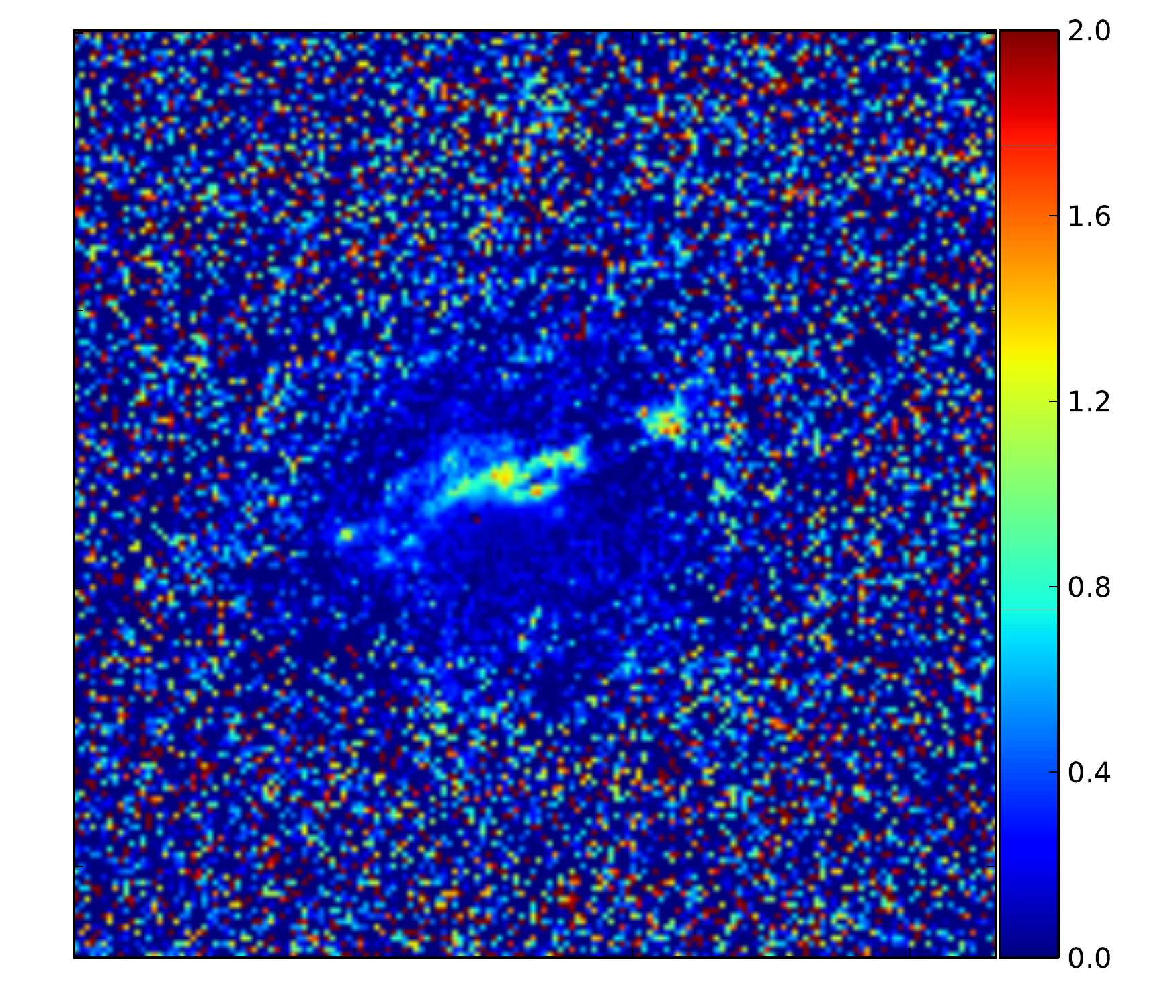}
\label{f:s3ds2_1482}}\\
\subfigure[\siii/\Ha]{
\includegraphics[width=3in]{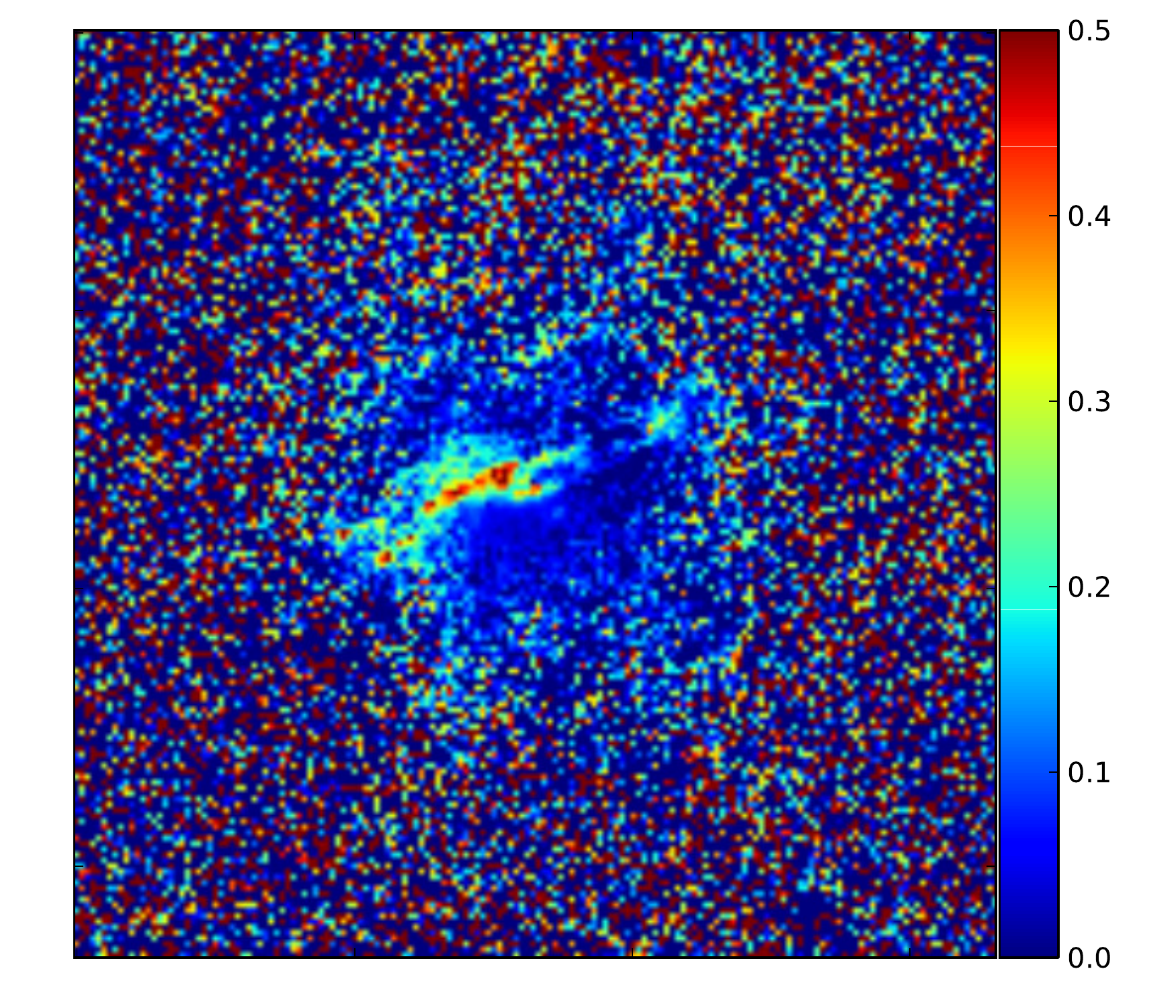}
\label{f:s3dha_1482}}\\
\subfigure[\sii/\Ha]{
\includegraphics[width=3in]{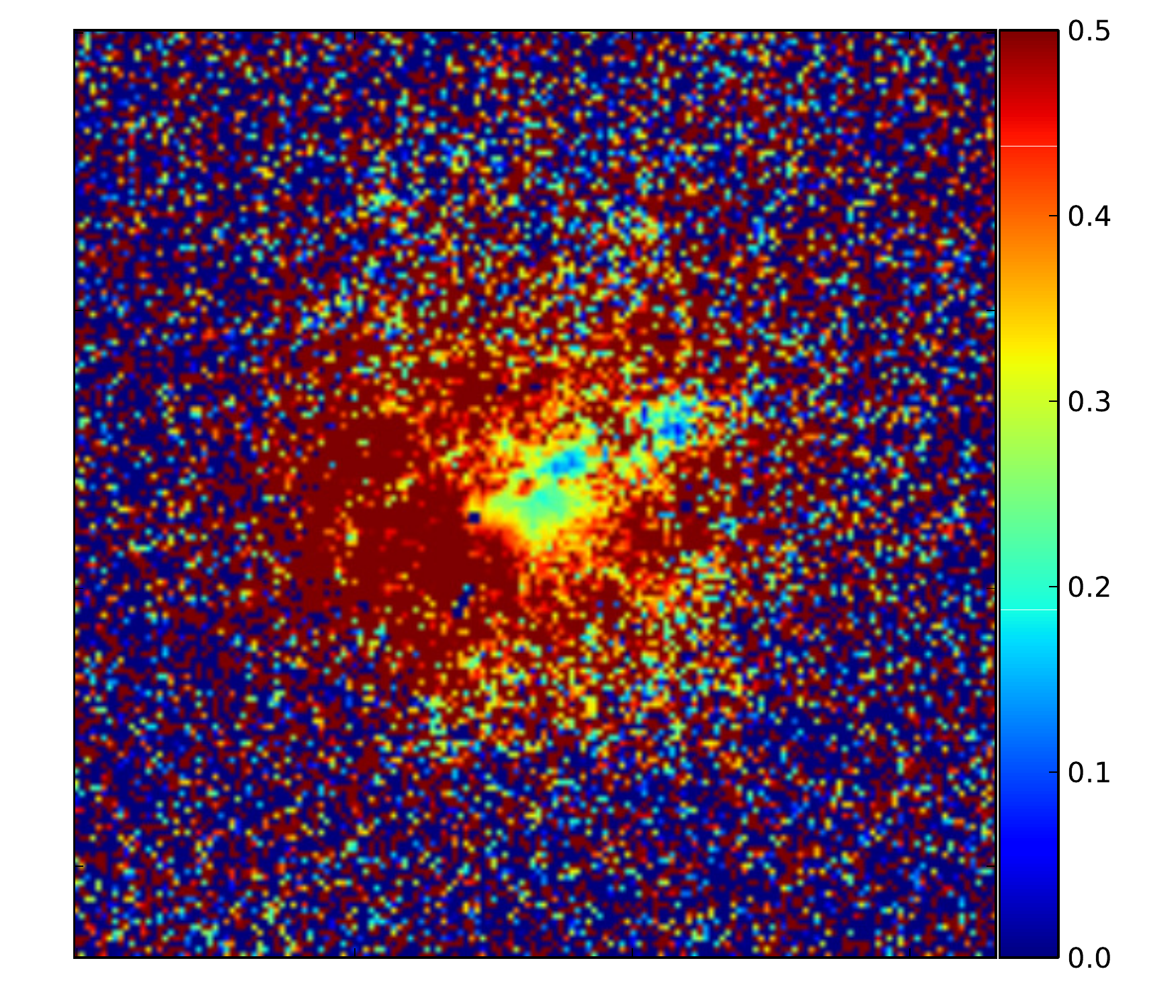}
\label{f:s2dha_1482}}
\caption{Emission-line ratio maps of for NGC~1482. These images have been binned 3x3, resulting in 0.6\arcsec\ pixel sizes. The maps are 41.67\arcmin\  on a side with an orientation such that North is up and East is to the left.}
\label{f:ratios_1482}
\end{figure}

In our emission-line images (Figure \ref{f:img_n1482}) and ratio maps (Figure \ref{f:ratios_1482}), we primarily detect the wind in \sii\ and \Ha.  The \siii\ emission is confined to the star forming region in the inner disk.   The ionization parameter map shows strong evidence for radiation bounding in NGC~1482. High \siii/\sii\ traces the disk star forming region, and the line ratio drops precipitously as one moves further into the wind (Figure \ref{f:s3ds2_1482}).  \citet{b:Veilleux_aj03} found the wind to be dominated by shock ionization based on \nii/\Ha\ observations. Our \sii/\Ha\ map (Figure \ref{f:s2dha_1482}) confirms this picture with \sii/\Ha\ $\geq 0.50$ in the wind. 

In reality, the emission observed in the wind will be a combination of shock- and photoionized gas.  In order to put strong constraints on the \siii/\sii\ line ratio in the wind, we explore correcting the line ratios for the shock ionization as follows. As discussed by \citet{b:Jaskot_apj13}, the observed line ratio can be written as:
\begin{equation}
\frac{A_o}{B_o} = \frac{A_s + A_p}{B_s + B_p}\label{eq:photo}
\end{equation}
where A and B correspond to the emission lines in question, and the subscripts $o$, $s$, and $p$ indicate the observed, shocked, and photoionized line strengths, respectively.  To convert our observed line ratios to the line ratios of photoionized gas, we need to solve for the fraction of shock-ionized relative to photoionized gas, 
\begin{equation}
X = \frac{H_s}{H_p}.
\end{equation}
Together with Equation~\ref{eq:photo}, this yields:
\begin{equation}
\frac{A_p}{B_p} = \frac{A_o}{B_o}  + X\left(\frac{A_o}{B_o} - \frac{A_s}{B_s}\right) \label{eq:step3}
\end{equation}.

\begin{figure}
\includegraphics[width=4in]{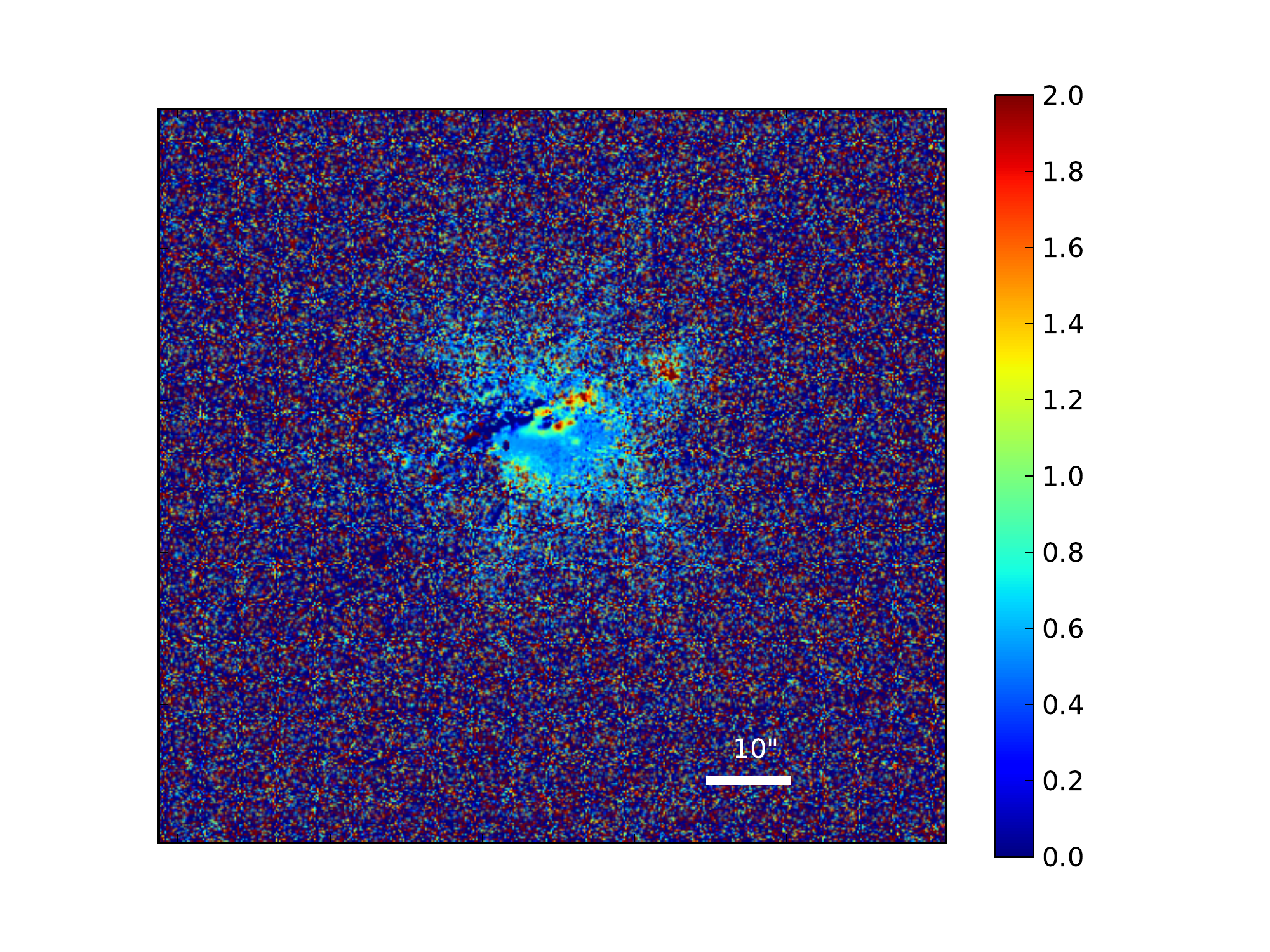}
\caption{Shock-corrected ionization parameter map of \siii/\sii.  This map is created assuming shock velocities of 950 km/s and the MAPPINGS III shock-only model \citep{b:Allen_apjs08}.  Even in the most optimistic correction, the \siii/\sii\ ratio transitions to values $<0.70$ close to the disk of the galaxy. N is up and E is to the left. \label{f:n1482_shock}}
\end{figure}

In Equation \ref{eq:step3}, $\frac{A_s}{B_s}$ can be replaced by the predicted line strengths from MAPPINGS III shock models \citep{b:Allen_apjs08}.   Since \citet{b:Veilleux_apj02} observed velocities of up to $v = 250$ km/s in the wind, we explore shocks with $v \leq 950$ km/s (assuming strong shocks).  For this calculation we use the MAPPINGS III shock-only models, which provide the most optimistic correction factor for high \siii/\sii\ from photoionization.  The precursor component to the shock would increase the relative contribution of shocks to the \siii, thus moving the corrected line ratios more towards optically thick than the shock only models. To solve for $X$, we write Equation \ref{eq:step3} for each emission line ratio, \siii/\sii, \siii/\Ha, and \sii/\Ha. By combining the resultant equations, we finally solve for $X$:
\begin{equation}
X = \rm \frac{\it{D_1}\left(\frac{[SII]}{\Ha}_o\right) + \it{D_2}\left(\frac{[SIII]}{[SII]}_o\right) + \it{D_3}}{\it{D_3D_2}}
\end{equation}
where
\begin{equation}
D_1 = \rm \frac{[SIII]}{[SII]}_o - \frac{[SIII]}{[SII]}_s\\
\end{equation}
and $D_2$ and $D_3$ are the corresponding values for \sii/\Ha\ and \siii/\Ha, respectively. After solving for $X$, we use that value in Equation \ref{eq:step3} to create a new \siii/\sii\ map (Figure \ref{f:n1482_shock}).  We find that the \siii/\sii\ maps remain strongly radiation bounded.   

Since there is evidence that the wind has cleared gas from the ISM of NGC~1482, it is somewhat surprising that this object does not show evidence for an optically thin wind. \HI\ observations reveal a hole in the \HI\ distribution at the location of the wind \citep{b:Hota_mnras05,b:Omar_japa05}.  Further evidence that the wind has cleared material from the center of the galaxy is found by \citet{b:Sharp_apj10}.  They generate a \sii$\lambda6716/\lambda6732$ ratio map and see a sharp gradient at the base of the wind, which suggests clearing of the inner disk by the wind \citep{b:Sharp_apj10}.  Both of these lines of evidence suggest that ionizing radiation can escape along the wind.  Perhaps the massive star population in NGC~1482 has aged in the time since the wind launched, and, thus, $Q_0$ has dropped significantly. Another explanation could be that dust entrained in the wind \citep{b:Vagshette_NewA12} provides additional material that absorbs radiation as it travels away from the galaxy.  

We note here that some of the emission may be shifted out of our very narrow bandpasses. For $v\sim250$\ km/s, the emission will shift by 5\AA\ and 7\AA\ for \sii\ and \siii, respectively.  This is comparable to half the bandpass in each of those filters.  Exacerbating this effect, our central bandpass in \siii\ is shifted $\sim4\AA$\ from the appropriate central bandpass for the redshift of NGC~1482.  This means that for NGC~1482 we do not detect some of the redshifted emission from \siii\ in the wind, although we should detect all of the blueshifted component. That being said, the wind exhibits the transition from high to low ionization parameter gas relatively close to the disk of the galaxy.  Furthermore, the \siii/\sii\ line ratio is well below 0.5 for most of the wind. Even if we assume that we are missing half of the \siii\ flux in the wind because of the velocity offset, the difference does not push the \siii/\sii\ ratio high enough to match our expectations for optically thin emission.  Given the 5 kpc extent of the wind \citep{b:Strickland_apjs04}, the location of the transition zone makes it less likely that the transition is a spurious one.  

\subsection{Non-Detections}\label{s:nondetect}

\subsubsection{NGC 178}

\begin{figure}[h]
\centering
\includegraphics[width=3.45in]{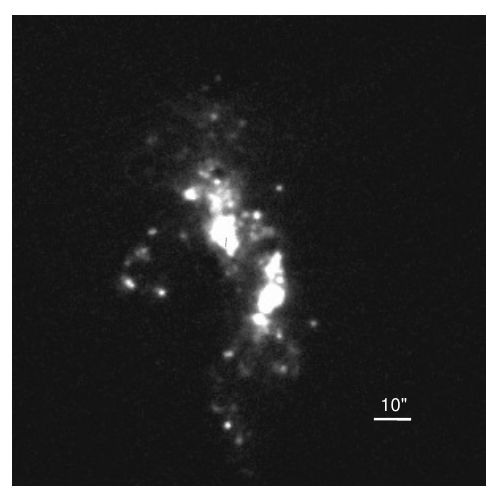}
\includegraphics[width=3.25in]{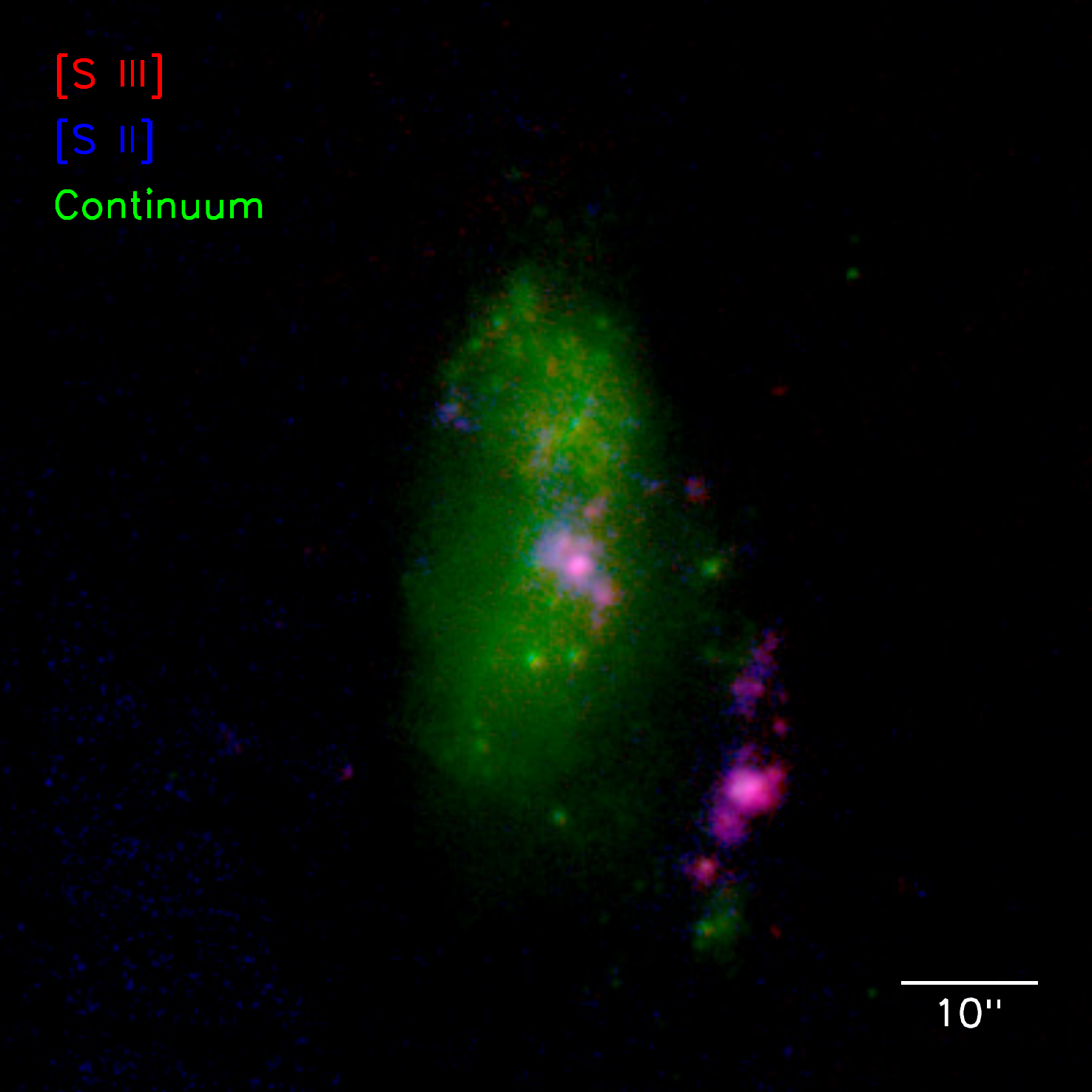}
\caption{Emission-line images of NGC 178. Top: \Ha\ Bottom: Three-color composite with rest frame \siii$\lambda9069$, \sii$\lambda6716$, and continuum at $\lambda6680$ in red, blue, and green, respectively. At 20.6 Mpc, 10\arcsec = 1 kpc. In this figure, N is up and E is to the left. \label{f:img_n178}}
\end{figure}

NGC 178 is forming stars at a rate of 0.55 \Msun/yr \citep{b:Oey_apj07} and lies 20.6 Mpc away \citep{b:Meurer_apjs06}.  In the emission-line images we see that ionized gas is split between two areas, one around the bulk of the observed continuum, and a separate extra-planar region (Figure \ref{f:img_n178}). Since ionizing radiation emitted at the edge of galaxies is more likely to escape \citep{b:Gnedin_apj08}, it might be reasonable to expect to find some evidence of density bounding in this region of the ionization parameter map.

\begin{figure}[h]
\subfigure[\siii/\sii]{
\includegraphics[width=3in]{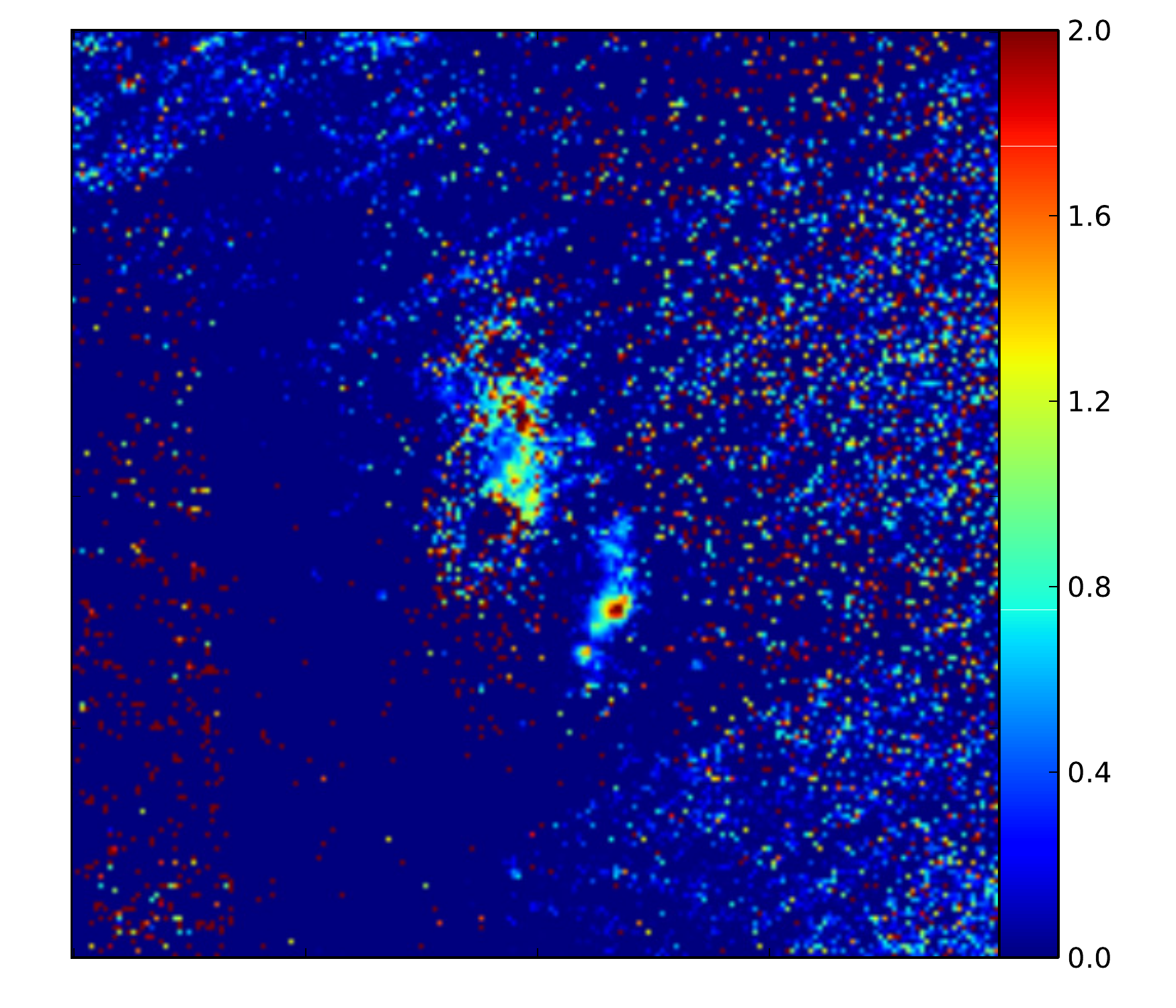}
\label{f:s3ds2_178}}\\
\subfigure[\siii/\Ha]{
\includegraphics[width=3in]{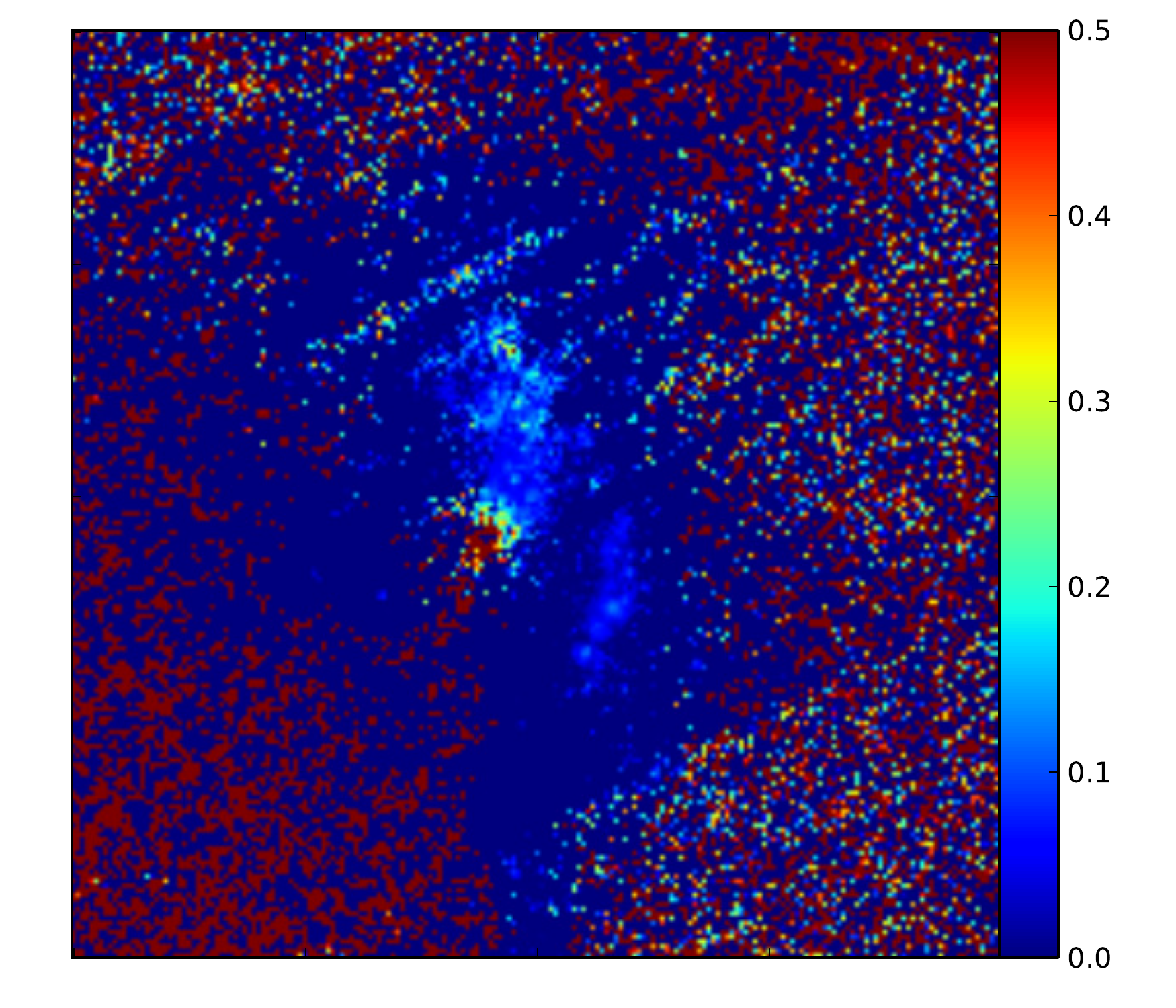}
\label{f:s3dha_178}}\\
\subfigure[\sii/\Ha]{
\includegraphics[width=3in]{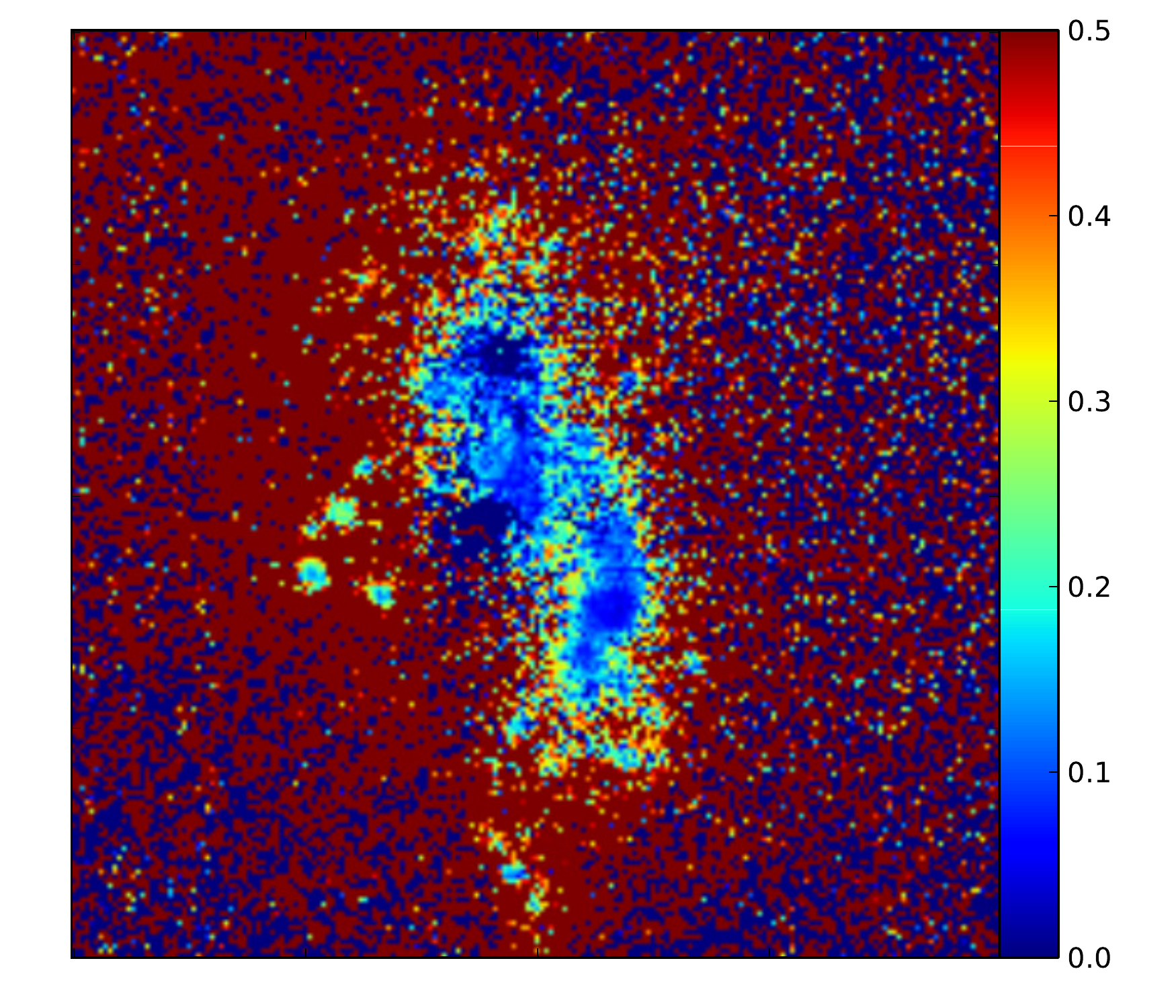}
\label{f:s2dha_178}}
\caption{Emission-line ratio maps of for NGC~178.  These images have been binned 3x3, resulting in 0.6\arcsec\ pixel sizes. The maps are 50\arcmin\  on a side with an orientation such that North is up and East is to the left.}
\label{f:ratios_178}
\end{figure}

However, the ionization parameter map shows that all nebular regions are optically thick.  The ionized gas to the west is comprised of a few separate peaks in the \siii/\sii\ ratio map (Figure \ref{f:ratios_178}), all of which exhibit a clear transition to low ionization parameter gas at the edges. 

The \sii/\Ha\ map (Figure \ref{f:ratios_178}) is intriguing. There appears to be an envelope of high \sii/\Ha\ surrounding the galaxy. This envelope could be caused by the well known \sii/\Ha\ gradient in the warm, ionized medium of galaxies \citep{b:Rand_apj98} or could be a signature of shocked gas surrounding the galaxy.  We note, however, that there are significant sky gradients and background features around the galaxy that are caused by scattered light within the instrument and by bright time observing conditions (\S \ref{s:data}). These sky features make it challenging to determine the nature of the surrounding low ionization parameter envelope. 

\subsubsection{NGC 7126}

\begin{figure}[h]
\centering
\includegraphics[width=3.45in]{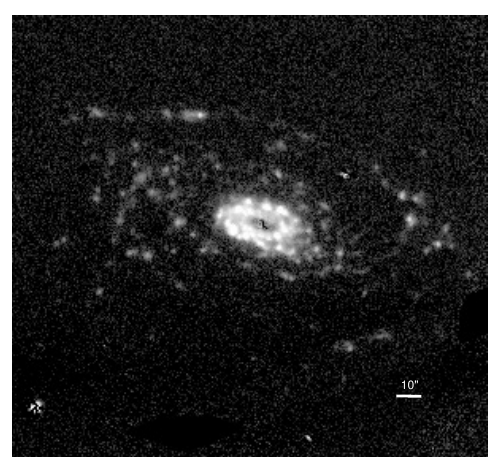}
\includegraphics[width=3.25in]{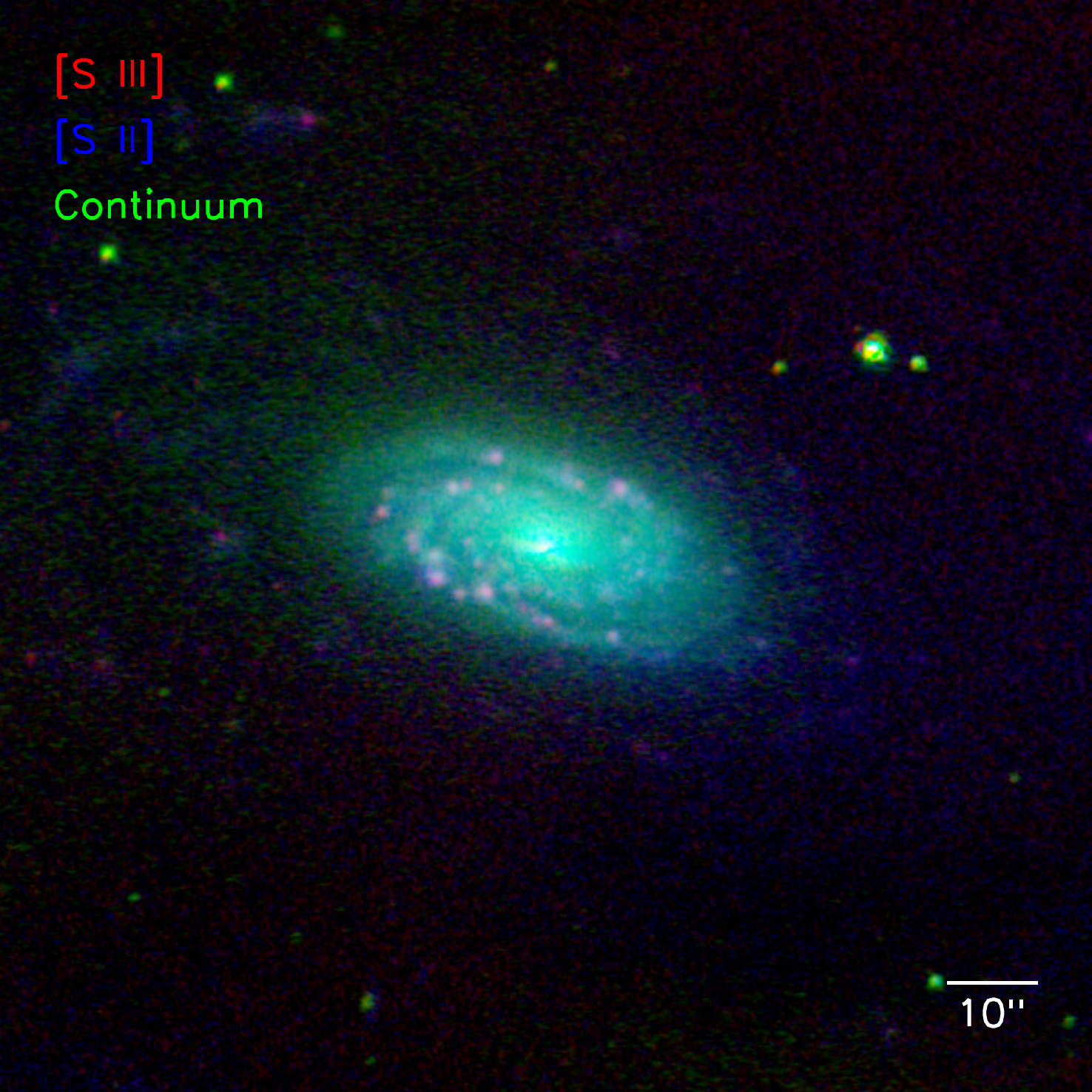}
\caption{Emission-line images of NGC 7126. Top: \Ha\ Bottom: Three-color composite with rest frame \siii$\lambda9069$, \sii$\lambda6716$, and continuum at $\lambda6680$ in red, blue, and green, respectively. At 45.5 Mpc, 10\arcsec = 2.3 kpc. In this figure, N is up and E is to the left. \label{f:img_n7126}}
\end{figure}

NGC 7126 is an SA(rs)c galaxy \citep{b:deVaucouleurs_rc391}. At 45.5 Mpc away \citep{b:Meurer_apjs06}, it is the most distant galaxy in our sample. The emission-line images reveal an extended spiral structure filled with \hii\ regions (Figure \ref{f:img_n7126}).  \HI\ observations show that it is interacting with its nearby companion NGC~7125 \citep{b:Nordgren_aj97}. NGC~7126 does not have a catalogued wind, and, with its high star formation rate, acts as a
control against this parameter in the sample \citep{b:Hanish_apj10}.  

\begin{figure}[h]
\subfigure[\siii/\sii]{
\includegraphics[width=3in]{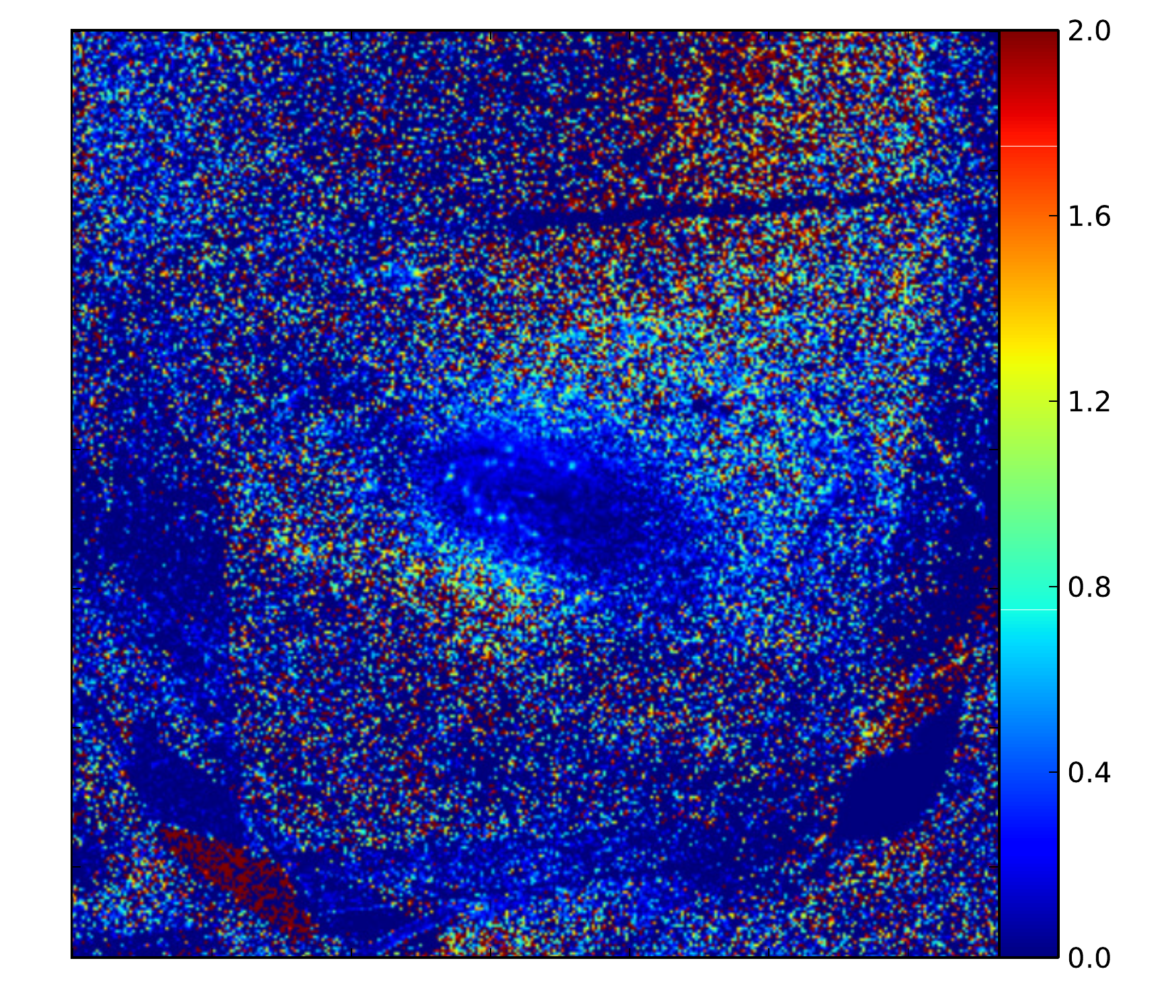}
\label{f:s3ds2_7126}}\\
\subfigure[\siii/\Ha]{
\includegraphics[width=3in]{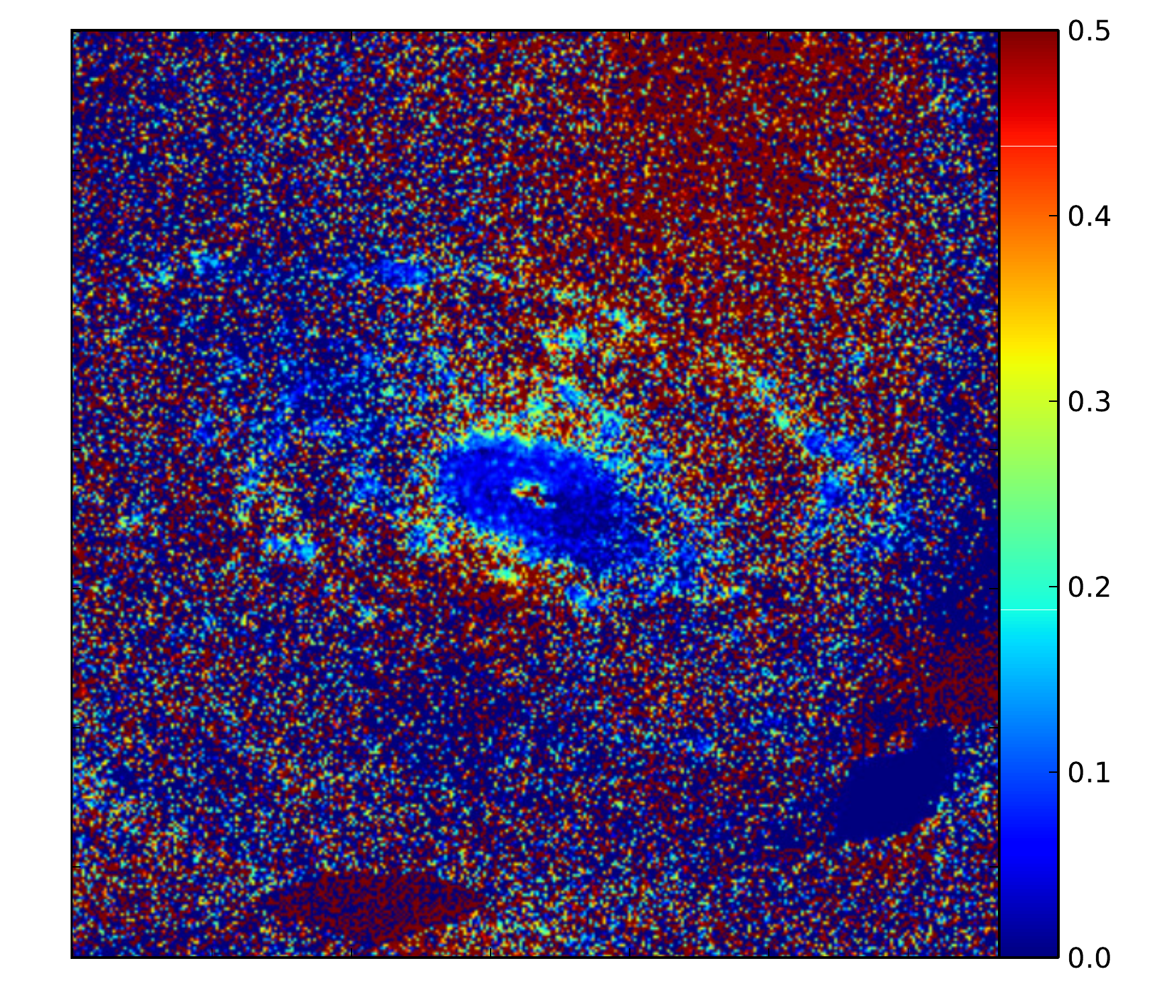}
\label{f:s3dha_7126}}\\
\subfigure[\sii/\Ha]{
\includegraphics[width=3in]{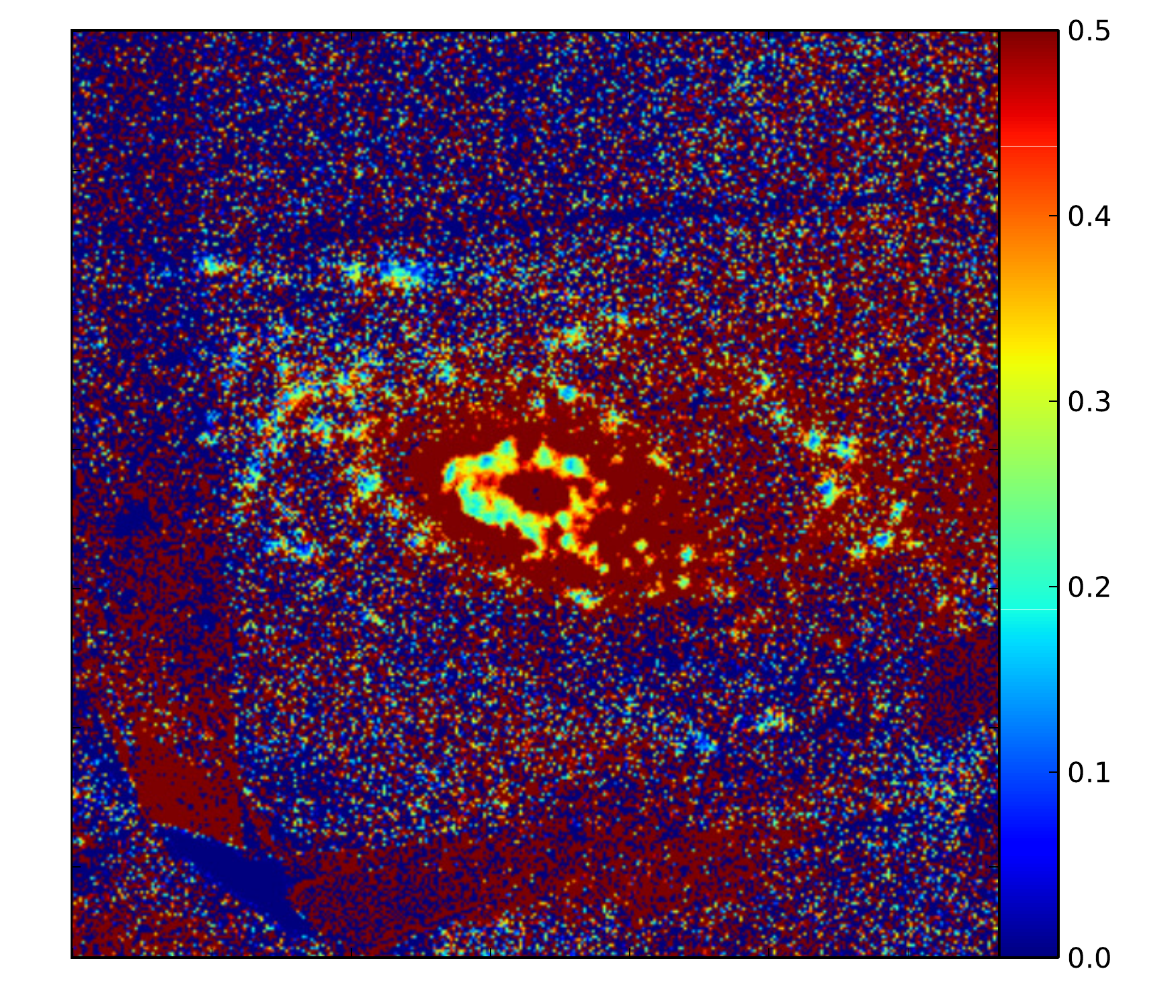}
\label{f:s2dha_7126}}
\caption{Emission-line ratio maps of for NGC~7126. These images have been binned 3x3, resulting in 0.6\arcsec\ pixel sizes. The maps are 83.33\arcmin\  on a side with an orientation such that North is up and East is to the left.}
\label{f:ratios_7126}
\end{figure}

The emission-line ratio maps are shown in Figure \ref{f:ratios_7126}.  The \siii/\sii\ map shows a remarkably low ratio of \siii/\sii. While the spiral structure is detected, it does not show up as clearly as it does in the \Ha\ ratio maps. In the \sii/\Ha\ ratio map, the \hii\ regions themselves appear with line ratios consistent with photoionized gas. As one moves to the edges of the individual regions, \sii/\Ha\ increases. This is consistent with expectations of the warm, ionized medium in galaxies \citep{b:Rand_apj98}.   We note, however, that these line ratios are a bit high, and they approach values consistent with shock ionization. The \siii/\Ha\ map exhibits relatively low line ratios throughout, even at the centers of the observed \hii\ regions. This could be indicative that the ionizing stellar population is composed primarily of B or late O type stars.

\section{Discussion}\label{s:discuss}

The full sample of seven galaxies provides examples of both optically thin and optically thick extended emission-line gas.  By comparing the properties of these galaxies we can gain insight on the processes and properties that influence the escape of ionizing radiation. Table \ref{t:SFR} presents relevant properties for each galaxy in the sample. Columns 1 and 2 are the galaxy name and \HI\ mass, respectively; Columns 3, 4, and 5 present the \Ha\ derived SFR, SFR surface density (\SigSFR), and the specific SFR (sSFR), respectively.  Column 6 provides the references for the star formation rates, \HI\ masses and \Ha\ radii. Finally, Column 7 indicates whether the galaxy shows wind and/or expanding bubble activity, and Column 8 shows the ratio of \HI\ mass to stellar mass (Table \ref{t:props}).

\begin{deluxetable*}{l c c c c c c c}
\tablewidth{0pt} 
\tabletypesize{\footnotesize}
\tablecaption{Star Formation Rates \label{t:SFR}}
\tablehead{\colhead{Galaxy}  & \colhead{log($M_{\HI}$)} & \colhead{SFR} & \colhead{\SigSFR$^a$} & \colhead{sSFR} &\colhead{Ref$^b$}  & \colhead{Wind or} & \colhead{M$_{\HI}$/M$_*^c$} \\ %
	\colhead{} & \colhead{[\Msun]} &\colhead{[\Msun/yr]}  &\colhead{[$\frac{\Msun}{\rm yr\ kpc^2}$]} & \colhead{Gyr$^{-1}$} & \colhead{} & \colhead{Bubbles?}&\colhead{} } %
\startdata 
\multicolumn{7}{c}{Optically thin extended emission}\\  
\hline\\
NGC~3125 &\nodata&0.461 &0.0826& 0.374& (1),n/a, (2)  & Y (3) &\nodata\\ 
NGC~5253 &8.22 & 0.428& 1.04  & 0.382& (1),(4),(5) & Y (3)  &0.006\\ 
\hline\\
\multicolumn{7}{c}{Optically thick extended emission}\\
\hline\\
NGC~1482 & 9.13&3.39 & 0.0271& 0.093& (1),(6),(7) & Y (8) & 0.037 \\  
NGC~1705 & 7.96  &0.057& 0.0091& 0.241& (1),(4),(9,7)& Y (10) & 0.389  \\ 
He~2-10  &8.49&0.872 &0.340 & 0.276 & (1),(11),(12) & Y (13) &0.098\\ 
\hline\\
\multicolumn{7}{c}{No detected extended emission}\\
\hline\\
NGC~178 & 9.40 &0.549 &0.019 & 0.315& (14),(4),(4) & N & 1.445 \\  
NGC~7126 & 10.57$^d$  &5.21 & 0.0467& 0.188& (14),(4),(4) & N &1.349  
\enddata
\tablenotetext{a}{\SigSFR\ is calculated from the \Ha\ SFR using either the radius of the ionized emission \citep{b:Calzetti_apj10} or the effective \Ha\ radius \citep{b:Oey_apj07}.}
\tablenotetext{b}{References, the first reference listed in the column is for the \Ha\ SFR, the second is for M$_{\HI}$, and the third is for $R_{\Ha}$ -- (1) \citet{b:Calzetti_apj10}; (2) \citet{b:GildePaz_apjs03}; (3) \citet{b:Marlowe_apj95}; (4) \citet{b:Meurer_apjs06}; (5) \citet{b:Calzetti_aj99}; (6) \citet{b:Hota_mnras05}; (7) \citet{b:Kennicutt_apj09}; (8) \citet{b:Veilleux_apj02}; (9) \citet{b:Kennicutt_apjs08}; (10) \citet{b:Meurer_aj92}; (11) \citet{b:Sauvage_aap97}; (12) \citet{b:Johnson_aj00}; (13) \citet{b:Mendez_aap99}; (17) \citet{b:Oey_apj07}}
\tablenotetext{c}{$M_*$ from Table \ref{t:props}.}
\tablenotetext{d}{Log($M_{\HI}$) is a combination of NGC 7126 and NGC 7125.}
\end{deluxetable*}

\subsection{Orientation bias}
The interpretation of escaping Lyman continuum observations in starbursts and LBGs is challenging because of the need to disentangle observational effects from intrinsic trends. One issue under debate is whether the relatively few ($\sim10\%$) galaxies that have significant \fesc\ are representative of the fraction of galaxies that have escaping Lyman continuum.  Potential challenges that can modify the intrinsic distribution of \fesc\ include low redshift interlopers, a preferred direction of escape, and small covering fractions for the escaping Lyman continuum \citep[e.g.,][]{b:Vanzella_apj12,b:Nestor_apj11}. Examining the morphologies of the ionization cones observed in this work can shed light on this issue.

Two of the seven starburst galaxies studied here show strong evidence for optically thin ionization cones.  In both NGC 5253 \citep[see also][]{b:Zastrow_apj11} and NGC 3125, the cones are aligned along the minor axis. This behavior is consistent with the expectations of massive-star feedback-driven models \citep[e.g.,][]{b:Murray_apj11,b:Kim_arXiv12}.  The observed alignment implies a preferred direction; feedback will punch holes in the ISM more easily along the minor axis \citep[e.g.,][]{b:Veilleux_araa05}. In a study of low redshift LBG analogs, \citet{b:Heckman_apj11} measured extreme wind velocities ($v\sim1500\ \rm km/s$) in all galaxies with low optical depth to Lyman continuum.  The high velocities indicate the direction of the wind must be nearly aligned with the line of sight to the galaxy.  This relationship between low optical depth and high outflow velocities further strengthens the argument that Lyman continuum escape has a preferred direction \citep{b:Zastrow_apj11}.  Furthermore, the ionization cones exhibit narrow morphologies and subtend an estimated 4\% of $4\pi$\ steradians.  This implies that the covering fraction of escaping radiation will be much less than unity, and is consistent with the results from studies of high redshift star-forming galaxies \citep[e.g.,][]{b:Nestor_apj11,b:Kreimeyer_arXiv13}.  We note that the narrow ionization cones do not necessarily imply small \fesc.  If the physical size of the cone's base is significantly larger than the size of the ionizing cluster, then large fractions of the ionizing flux produced by the cluster may still escape through the angularly narrow cone.  The combination of small opening angles and preferred directions suggest that if one tried to directly detect escaping Lyman continuum, the axis of escape would need to be close to the observer's line of sight in order to obtain a strong detection.  If more galaxies are like NGC 3125 and NGC 5253, then an orientation bias makes it more challenging to detect escaping Lyman continuum in starbursts and LBGs.

\subsection{Possible Trends with Galaxy Properties}

In light of the massive-star feedback-driven model discussed previously, the first galactic property to examine is the presence of a galactic wind.  All of the galaxies that show evidence for extended emission also have evidence for either a galactic wind or superbubble expansion out of the galaxy.  Thus confirming the link between winds and the fate of ionizing radiation \citep[e.g.,][]{b:Heckman_apj11}.  However, NGC~1705 and NGC~1482, the two galaxies that have the most clearly established winds, both appear optically thick.  We note that both galaxies without extended emission, NGC 178 and NGC 7126, do not have galactic winds. Thus, as discussed in \citet{b:Heckman_aj97}, we see that while wind activity is important for the escape of Lyman continuum, other factors clearly play a determining role.

Another critical property to examine is star formation rate.  Models predict that escaping ionizing radiation will be associated with extreme feedback from strongly star forming systems \citep[e.g.,][]{b:Clarke_mnras02,b:Wise_apj09,b:Paardekooper_aap11}.  From Table \ref{t:SFR}, it is immediately apparent that SFR(\Ha) is not directly correlated with the optical depth of extended emission. The galaxy with the highest SFR, NGC 7126, is one that most clearly does not have extended optically thin emission. Furthermore, the galaxy with the second highest SFR, NGC 1482, is optically thick. 

A more meaningful comparison is to look at the \SigSFR\ and the specific SFR.  These measures of the SFR take into account the sizes and masses of the galaxies.  A more massive galaxy would need a higher SFR to produce enough feedback for ionizing radiation to escape.  Table \ref{t:SFR} shows that NGC~178 and NGC~7126, which show no evidence for extended emission, do have low \SigSFR, which is consistent with our expectations. However, distinguishing between the galaxies that have optically thin and thick emission is more challenging. NGC~5253 has the highest \SigSFR, but NGG~3125 has \SigSFR\ lower than He~2-10, which is likely optically thick.  One explanation is that the higher SFR in He~2-10 contains contamination from its AGN. However, the AGN in He~2-10 is not particularly luminous in \Ha\ \citep{b:Reines_nature11} making the significance of its contribution unclear.  NGC~1705 has comparatively low \SigSFR. Paired with the lower \SigSFR\ of NGC~1482, the rates are consistent with the hypothesis that optical depth is correlated with \SigSFR.   However, we cannot extrapolate any conclusions on trends with \SigSFR\, based only on our findings for these individual galaxies. 

The comparison between sSFR is similarly fraught.  NGC~5253 and NGC~3125 do have among the highest sSFR, but their sSFRs are not significantly higher than those of the galaxies with optically thick emission.  Furthermore, NGC~178, which doesn't have clear evidence for extended emission, has a rather high sSFR. Thus, our small sample supports previous work in its conclusion that strong SFR is necessary but not sufficient for significant \fesc\ \citep{b:Heckman_apj01}.  We caution the reader that our sample is small, and the range of star-formation rates is also small.  Therefore, our sample does not provide the leverage to firmly establish any apparent correlation between optical depth and \SigSFR\ or sSFR. As an additional consequence of the small sample size, the galaxies that seem anomalous, such as NGC 178 with its high sSFR but lack of extended emission and He~2-10, which is optically thick despite its high \SigSFR, may not be significant outliers.

In addition to SFR, the recent star formation history will have an influential role on the escape of ionizing radiation.  As a stellar population ages, the rate of ionizing photons produced (\Qo) decreases and the hardness of the radiation field softens.  The ages of the clusters responsible for the ionization cones in both NGC~5253 and NGC~3125 are between 2-5 Myr old \citep{b:Calzetti_aj97,b:Westera_aap04,b:Chandar_apj05,b:Raimann_mnras00}.  In contrast with this, NGC~1705 is in a post-burst state (cluster ages $\sim 10-15$ Myr) in which most of the O stars in the main super star clusters have died off \citep{b:Heckman_aj97}.  Thus, the softer and fainter B stars dominate the radiation field and do not provide enough ionizing radiation to allow for escape.  This may similarly offer a partial explanation for the line ratios observed in He~2-10.  While the main star forming region, embedded in the center of the galaxy, contains many young clusters with ages ranging from 2-20 Myr \citep{b:Johnson_aj00}, the second star-forming region, further towards the eastern edge of the galaxy is likely much older with fewer of its massive stars remaining.  Furthermore, visual examination of the images of He~2-10 show closed bubbles, rather than fractured filaments (Figure \ref{f:img_he210}).  We note that for radiation fields dominated by late type stars, the relationship between the ionization parameter map and the optical depth is complicated.  In these situations, a transition to \sii\ dominated gas does not necessarily indicate an optically thick nebula \citep{b:Pellegrini_apj12}.  If the radiation field is soft but sufficiently strong, it is possible that not enough gas is present to absorb the lower energy photons, yet, the ionization parameter map still exhibits a transition zone to \sii. 

Furthermore, whether the population is a single burst or an extended burst also influences the fate of ionizing radiation.  While feedback generates low density passageways in the ISM, simulations of superbubble evolution show that there is a delay between superbubble formation and the actual escape of ionizing radiation  \citep{b:Dove_apj00,b:Fujita_apj03}.  During the phase prior to superbubble break-out, ionizing radiation gets trapped by the bubble walls \citep{b:Fujita_apj03}.  Once the bubble breaks out of the disk, ionizing radiation may escape relatively unhindered. However, by this point in time, the rate of ionizing photons produced by the massive star population is considerably weaker than in the first few Myr \citep{b:Dove_apj00,b:Fujita_apj03}.  If the starburst occurs over a short duration, essentially a single burst, then it is less likely to leak significant ionizing radiation.  If instead the burst is more extended, the later episodes of star formation can capitalize on the passageways carved by prior star forming episodes. Observationally, \citet{b:McQuinn_apj10b} showed that extended bursts with typical timescales of hundreds of Myr are common in dwarf starbursts.

Our data further support this scenario.  In NGC~5253, the current
burst of star formation occurred within the last 3--5 Myr
\citep{b:Calzetti_aj97}. While it is unlikely that any superbubbles
formed from this epoch will have broken from the disk, there is
evidence for prior star formation episodes with ages on the order of
10--100 Myr \citep{b:Caldwell_apj89}.
Thus, the ionization cone in NGC~5253 probably formed through low
density regions cleared out by feedback from earlier episodes of star
formation.  In NGC~3125, the super star cluster at the base of the
ionization cone has an age between 1--3 Myr, while the region
surrounding it has an overall mean age of between 8--10 Myr
\citep{b:Westera_aap04,b:Chandar_apj05}. While the initial episode of
star formation in this burst is younger than that shown for NGC~5253,
the dominant star formation is on the periphery of the galaxy, which
makes it easier for ionizing radiation to escape
\citep{b:Gnedin_apj08}.  Furthermore, \citet{b:Raimann_mnras00}
stacked a sample of similar HII galaxies in which NGC~3125 was the
dominant source and found that 65\% of the stellar population is
younger than 5 Myr, 32\% is between 5 and 100 Myr, and the remaining
3\% of the population us is older than 100 Myr.  While the SFH
analysis is not based solely on NGC~3125 it does point in the
direction of NGC~3125 having multiple recent episodes of star
formation. 

The young ages of the ionizing populations in NGC~3125 and NGC~5253 may represent a special age range needed to find escaping ionizing radiation.  In a recent paper, \citet{b:Jaskot_apj13} study a sample of $z\sim0.1-0.3$ galaxies whose extremely high \oiii/\oii\ suggest that they are likely optically thin.  These galaxies all have ionizing populations that are between 3-5 Myr \citep{b:Jaskot_apj13}, similar to the ages observed in galaxies with ionization cones studied in this work. Furthermore, the two local galaxies that have direct \fesc\ measurements, both have significant populations of young stars \citep{b:Leitet_aap13, b:Leitet_aap11}.     

Finally, it has been suggested that the mass of the galaxy is
correlated with the escape fraction of ionizing radiation.  There are
some simulations that predict higher \fesc\ with larger galaxy mass
\citep[e.g.,][]{b:Gnedin_apj08}.  Others claim the opposite because
the smaller galaxy's weaker potential well would make escape easier
\citep{b:Yajima_mnras11,b:Razoumov_apj10}.  Based on our small sample
here, we do not find evidence for or against either claim.  The two
galaxies for which the extended ionized gas is optically thin have
masses that are in the middle to low range of masses in the sample
(Tables \ref{t:props} and \ref{t:SFR}). Interesting to note, these two
galaxies have significantly higher fractions of \HI\ relative to the
stellar mass of the galaxy (Table \ref{t:SFR}) than the sample
galaxies with extended ionized gas. Meanwhile, NGC~5253 has an
exceedingly low \HI\ mass fraction, even when compared to the other
galaxies with winds. This is consistent with the idea that the
clearing, or consumption, of the neutral gas in the galaxy plays a
role in regulating the escape of ionizing radiation.

\section{Conclusions} \label{s:conc}

The passage of ionizing radiation through, and possibly out of, a galaxy has ramifications on our understanding of cosmic reionization.  In this paper, we study the extended, low surface brightness, emission-line gas of seven dwarf starburst galaxies.  Using narrowband \siii, \sii, and \Ha\ images and the technique of ionization parameter mapping, we evaluate the optical depth of the extended emission.  

In two of the galaxies, we discover optically thin ionization cones extending along the minor axis.  These cones suggest that ionizing radiation is escaping the main body of the galaxy.  The narrow morphology suggests that ISM morphology is a critical determinant of whether ionizing photons will escape a galaxy.  In three of the galaxies, we find that the established galactic winds are most likely optically thick.  These galaxies, despite the presence of strong feedback do not show clear evidence for escaping ionizing radiation.  The two remaining galaxies, despite the strong star formation show little evidence of extended emission.  

Our most convincing examples of optically thin emission are found in narrow ionization cones.  These narrow cones illustrate a scenario in which the covering fraction of Lyman continuum is significantly less than unity, as has been suggested by other work \citep[e.g.,][]{b:Nestor_apj11}. Furthermore, both cones are aligned with the minor axis of the galaxy, indicating a preferred direction.  The small opening angle and preferred direction suggest that unless the galaxy orientation is such that the axis of escape aligns with the line of sight, it will be challenging to directly detect escaping radiation. If other starbursts are similar to NGC~3125 and NGC~5253, an orientation bias at least partially explains the low detection rate of ionizing radiation in starburst galaxies and LBGs.

In addition to observational biases, we explore the galactic properties that may regulate the escape of ionizing radiation.  While galactic winds and bubble activity encourage escaping radiation, not all starbursts with galactic wind have escaping Lyman continuum.  One explanation is that by the time the bubble or wind breaks out of the ISM, the ionizing population has aged and is no longer producing ionizing radiation as prodigiously \citep[e.g.,][]{b:Dove_apj00,b:Fujita_apj03}.  Alternatively, the ionizing population is too young to have carved sufficient low-density paths out of the ISM for escape.  This suggests an optimal age between 3-5 Myr for the escape of ionizing radiation.  Based on the recent star-formation histories of NGC~3125 and NGC~5253, a starburst with a few recent star forming episodes strikes the balance between the time it takes for bubbles to burst and maximizing the number of ionizing photons produced that have clear passage out of the galaxy.  

Another potential contributing factor is the concentration of star formation in the burst. In the examples of optically thin ionization cones, we see that they emanate from specific, concentrated star forming regions, rather than extended star-forming areas.  In contrast, NGC~7126 shows no sign for feedback-driven extended emission, and has its copious star formation spread throughout the disk of the galaxy.  This suggests that concentrated star formation will be more effective at clearing out the ISM.  However, we must be careful in drawing strong conclusions from so few objects.  He~2-10 and NGG~1705, which also have concentrated star formation, are most likely optically thick to ionizing radiation. Thus, while the concentration of star formation is important, the interplay between the star formation and the ISM morphology most likely regulates the escape of ionizing radiation from starbursts and Lyman break galaxies. 

\acknowledgments

Support for this work was provided by the National Science Foundation,
grants AST-0806476, AST-0907758, and AST-1210285 to MSO; and
AST-0606932 and AST-1009583 to SV and MM.  We also acknowledge a Margaret  
and Herman Sokol Faculty Award to MSO.  MM acknowledges support by NASA through a Hubble Fellowship grant HST-HF51308.01-A awarded by the Space Telescope Science Institute, which is operated by the Association of Universities for Research in Astronomy, Inc., for NASA, under contract NAS 5-26555. We are grateful to Colin Slater for help collecting observations and the UM FANG research group, Eric Pellegrini, Amy Reines, Lee Hartmann, and Timothy McKay for helpful discussions. 


\clearpage

\end{document}